\documentclass[twocolumn,showpacs,aps,prx,superscriptaddress]{revtex4-1}
\usepackage[bottom]{footmisc}
\usepackage{color}
\usepackage[unicode=true,pdfusetitle,bookmarks=true,
bookmarksnumbered=false,bookmarksopen=false,breaklinks=true,pdfborder={0
0 0},backref=false,colorlinks=true]{hyperref}
\hypersetup{linkcolor=blue,citecolor=blue,urlcolor=blue} \usepackage{breakurl}
\usepackage{amssymb}
\usepackage{graphicx}
\usepackage{amsmath}
\usepackage{epstopdf}
\usepackage{tensor}
\usepackage{slashed}
\usepackage{cancel}

\usepackage{feynmp-auto}
\newcommand{\tr}{\text{tr}}

\newcommand{\nn}{\nonumber}
\renewcommand{\L}{\mathcal{L}}
\renewcommand{\b}[1]{{\boldsymbol{#1}}}
\renewcommand{\c}[1]{\mathcal{#1}}
\renewcommand{\geq}{\geqslant}

\def\beq#1\eeq{\begin{align}#1\end{align}}
\newcommand{\lb}{\nonumber\\}

\graphicspath{{./}{./figs/}}

\begin{document}

\title{Quantum-critical electrodynamics of Luttinger fermions}
\date{\today}

\author{Santanu Dey}
\affiliation{Department of Physics, University of Alberta, Edmonton, Alberta, T6G 2E1, Canada}
\author{Joseph Maciejko}
\affiliation{Department of Physics, University of Alberta, Edmonton, Alberta, T6G 2E1, Canada}
\affiliation{Theoretical Physics Institute (TPI), University of Alberta, Edmonton, Alberta, T6G 2E1, Canada}

\begin{abstract}
We study the quantum electrodynamics of Luttinger fermions with quadratic band-crossing dispersion in three dimensions. The model can be viewed as the low-energy effective theory of a putative $U(1)$ quantum spin liquid with fermionic Luttinger spinons, or as an extension of the Luttinger-Abrikosov-Beneslavskii (LAB) model that accounts for transverse gauge fluctuations with finite photon velocity. Aided by a renormalization group analysis below four dimensions, we elucidate the presence and stability of quantum critical phenomena in this model. We find that the non-Fermi liquid LAB phase is stable against gauge fluctuations, and can thus also be viewed as a $U(1)$ spin liquid with gapless Luttinger spinons. We discover a multicritical point with Lifshitz scaling that corresponds to a time-reversal symmetry-breaking quantum phase transition from the LAB state to a chiral spin liquid with spinon Landau levels and birefringent emergent photons. This multicritical point is characterized by a finite fermion-photon coupling in the infrared and can be viewed as a fermionic analog of the Rokhsar-Kivelson point in three-dimensional quantum dimer models.
\end{abstract}

\maketitle

\section{Introduction}
\label{intro}

Quantum field theories of gapless fermions coupled to dynamical $U(1)$ gauge fields play a key role in contemporary physics. In high-energy physics, relativistic quantum electrodynamics (QED) describes fundamental light-matter interactions and historically served as the first example of unification of quantum mechanics and special relativity. In condensed matter physics, fermionic gauge theories arise naturally as low-energy effective descriptions of strongly correlated systems~\cite{XGWen}. In frustrated magnetism, relativistic QED in 2+1 dimensions serves as an effective description of the algebraic or Dirac spin liquid, a stable critical state of interacting spins with universal power-law correlations~\cite{affleck1988,marston1989,kim1999,rantner2001,rantner2002,hermele2004,hermele2005}. Quantum phase transitions out of the Dirac spin liquid are described by the critical fixed points of various relativistic QED-Gross-Neveu theories, where gauge interactions are supplemented with four-fermion contact interactions~\cite{janssen2017,ihrig2018,zerf2018,boyack2018,boyack2019,zerf2019,dupuis2019,dupuis2019b,dupuis2021,
dupuis2021b,zerf2020,janssen2020,boyack2021,boyack2021b}.

In condensed matter systems, Lorentz invariance can emerge as a symmetry of the low-energy continuum field theory only if the fermion band structure exhibits symmetry-protected linear (Dirac/Weyl) crossings in the noninteracting limit. The more generic case is that of an extended Fermi surface coupled to a fluctuating gauge field, which appears in the description of $U(1)$ spin liquids with a spinon Fermi surface~\cite{baskaran1987,lee2005,lee2008,metlitski2010,mross10,holder2015} and the composite Fermi liquid in the half-filled Landau level~\cite{halperin93,nayak1994}. Such (2+1)-dimensional nonrelativistic gauge theories also model possible quantum phase transitions out of the spinon Fermi surface state or the composite Fermi liquid, e.g., pairing transitions~\cite{metlitski2015}.

In this work, we study a situation intermediate between the aforementioned examples of relativistic fermions and Fermi surfaces: Luttinger fermions coupled to a $U(1)$ gauge field in 3+1 dimensions. The nonrelativistic Luttinger model with dynamic critical exponent $z=2$ describes the electronic structure of a spin-orbit coupled material in the vicinity of a quadratic band crossing (QBC) protected by cubic point-group symmetries~\cite{luttinger56}. It has been applied to various compounds of recent interest such as inverted band gap semiconductors (HgTe, $\alpha$-Sn)~\cite{murakami04} and the pyrochlore iridates A$_2$Ir$_2$O$_7$, where A is a lanthanide element~\cite{moon13}. Owing to its gauge structure, our model can be viewed as the effective theory of a putative (3+1)-dimensional $U(1)$ spin liquid with fermionic Luttinger spinons. (Spin liquids with QBC spinons in 2+1 dimensions were proposed in Refs.~\cite{xu2012,mishmash2013}, and studies of $U(1)$ spin liquids in 3+1 dimensions have largely focused on the Coulomb spin liquid~\cite{huse2003,moessner03,hermele04,savary2012} with no gapless matter excitations.) To elucidate the low-energy physics of our gauge theory of Luttinger fermions, we employ a controlled perturbative renormalization group (RG) analysis in $d=4-\epsilon$ spatial dimensions, and search for fixed points of the RG flow. Stable fixed points correspond to stable phases of matter, i.e., spin liquid or broken-symmetry phases, while unstable fixed points correspond to possible (multi)critical points associated with phase transitions.

\begin{figure}[t]
  \includegraphics[width=\columnwidth]{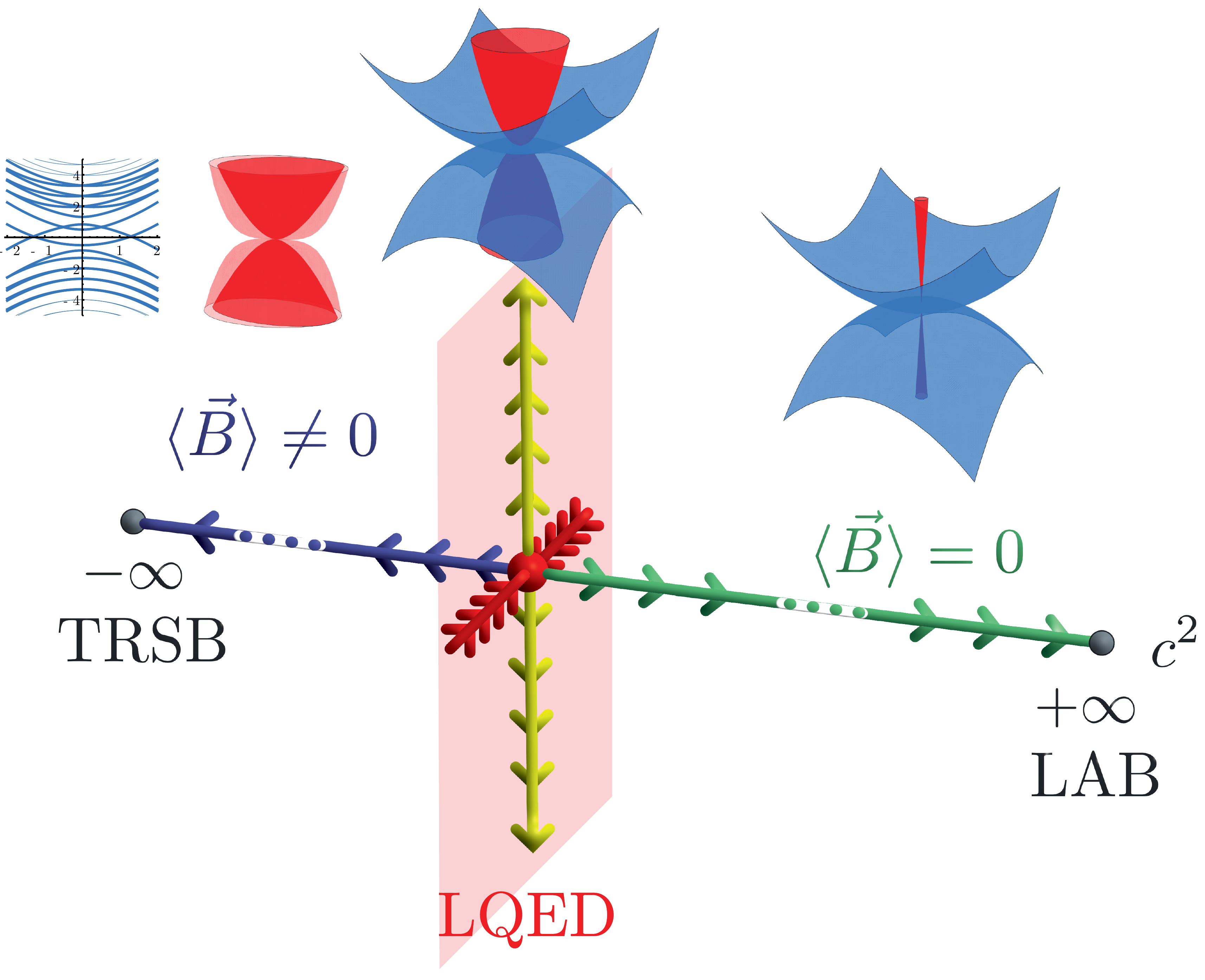}
   \caption{Quantum criticality of rotationally and particle-hole symmetric Luttinger fermions (blue dispersion) coupled to a dynamical $U(1)$ gauge field (red dispersion). The photon velocity squared $c^2$ is a relevant coupling; it drives a transition between a symmetric LAB phase with instantaneous photons and a time-reversal symmetry-breaking (TRSB) phase with a spontaneously generated background magnetic field, Landau levels for Luttinger fermions, and birefringent photons. For more than one fermion flavor, the phase boundary contains a critical line governed by a Lifshitz-QED (LQED) theory with finite fermion-photon coupling and anisotropic $z>2$ scaling for both fermions and photons.}
   \label{phase_diag}
\end{figure}

Our main results, which are illustrated in Fig.~\ref{phase_diag}, can be summarized as follows. Our model initially consists of rotationally invariant, particle-hole symmetric Luttinger fermions minimally coupled to Maxwell electrodynamics with dynamical electric and magnetic fields $\b{E},\b{B}$. Requiring $z=2$ scaling at tree level, we find that the theory admits an additional derivative coupling $(\nabla\times\b{B})^2$ in the gauge sector. In our RG treatment, the photon velocity squared, $c^2$, emerges as a relevant tuning parameter for quantum phase transitions. The only symmetric, perturbatively accessible, stable fixed point of the theory has $c^2_*=\infty$ and is equivalent to the Luttinger-Abrikosov-Beneslavskii (LAB) fixed point of electrons interacting via the static Coulomb interaction~\cite{moon13,abrikosov1971,abrikosov1974}. Our first main result is thus that, in the rotationally invariant and particle-hole symmetric limits, the LAB fixed point is stable against transverse gauge fluctuations for arbitrary number $N_f$ of fermion flavors. In the current context, it corresponds to a stable critical spin liquid in 3+1 dimensions with QBC spinons, analogous to the (2+1)-dimensional algebraic spin liquid. In addition to the LAB fixed point, for $N_f\geq 2$ we find an unstable fixed point with $c^2_*\sim\mathcal{O}(\epsilon)$. Since $c^2$ is a relevant parameter, we interpret it as a Wilson-Fisher-like multicritical point governing a quantum phase transition between the LAB phase and a broken-symmetry phase at $c^2_*=-\infty$. The latter exhibits a spontaneously generated, time-reversal symmetry-breaking background magnetic field $\langle\b{B}\rangle\neq 0$ orbitally coupled to the Luttinger fermions: it thus corresponds to a type of chiral spin liquid with emergent, three-dimensional Landau levels. The gauge sector exhibits birefringence, i.e., an ``ordinary'' Lifshitz photon with isotropic $z=2$ dispersion, as well as an ``extraordinary'' photon with $z=2$ scaling along the background field and $z=1$ scaling perpendicular to the field. At criticality, the gauge sector is again described by a (spatially isotropic) quantum Lifshitz gauge theory with a nonrelativistic photon dispersion, $\omega\sim k^z$, which also describes the Rokhsar-Kivelson point of quantum dimer models in 3+1 dimensions~\cite{moessner03,hermele04}. To the difference of the latter, however, the Lifshitz electrodynamics at our multicritical point is strongly coupled to the fermion sector, leading to non-Gaussian corrections $z-2>0$ to the dynamic critical exponent. While Lorentz-violating gauge theories have been investigated in a high-energy physics context~\cite{das2009,anselmi09,anselmi10,farias12,farias2013,alexandre13,gomes2015}, we are not aware that interacting fixed points of gauged QBC fermions have been discussed previously. Quantum critical phenomena with Luttinger fermions have been discussed extensively~\cite{savary2014,herbut2014,janssen15,janssen16,boettcher16,janssen2017b,boettcher17}, but in the absence of propagating photons (static limit).

The rest of the paper is organized as follows. Sec.~\ref{model} describes a theory of rotationally invariant, particle-hole symmetric Luttinger fermions coupled to a dynamical $U(1)$ gauge field. Sec.~\ref{rg} details the Wilsonian RG analysis of the theory by means of an $\epsilon=4-d$ expansion near the upper-critical dimension $d=4$. Sec.~\ref{lab} discusses the first important result of our analysis: the theory admits a symmetric LAB phase that is stable against transverse gauge fluctuations. Sec.~\ref{SSB} focuses on the time-reversal symmetry-breaking phase, and Sec.~\ref{mcp} describes the interacting Lifshitz-QED (LQED) multicritical point that intervenes between the two stable phases. In Sec.~\ref{summary}, we summarize our findings and outline directions for future research.

\section{Model}
\label{model}

In this section, we describe the $U(1)$ gauge theory that dictates
the low-energy behavior of charged, QBC Luttinger fermions. The imaginary-time Lagrangian is:
\begin{align}\label{L}
\c{L}=\c{L}_\psi+\c{L}_A,
\end{align}
where $\c{L}_\psi$ describes the fermion sector and $\c{L}_A$ the gauge sector. The simplest QBC Hamiltonian in $d$ spatial dimensions can be written down in the form $H=-\sum_{i,j=1}^d G_{ij}p_ip_j$, where $G_{ij}$ transforms as a second-rank symmetric tensor under rotations. Considering for simplicity a particle-hole symmetric dispersion $E_{\pm}(\b{p})=\pm p^2/2m$ with effective mass $m$, the Lagrangian $\c{L}_\psi$ of the QBC fermions minimally coupled to a $U(1)$ gauge field can be expressed in
terms of $N_\Gamma = (d-1)(d+2)/2$ Hermitian gamma matrices $\Gamma^a$ obeying the Euclidean 
Clifford algebra $\{\Gamma^a,\Gamma^b\}=2\delta^{ab}$ (see App.~\ref{fermion}):
\begin{align}\label{Lpsi}
\c{L}_\psi=\sum_{\alpha=1}^{N_f}\psi_\alpha^\dagger\left(D_\tau+\frac{\Gamma^a}{2m}d_a(-i\b{D})\right)\psi_\alpha,
\end{align}
where $\psi_\alpha$ denotes $N_f$ flavors of Luttinger fermions, which are $d_\Gamma$-dimensional spinors. Here $d_\Gamma=2^{\lfloor N_\Gamma/2\rfloor}$ where $\lfloor\cdot\rfloor$ is the floor function. $D_\mu=\partial_\mu+ieA_\mu$ denotes the gauge-covariant derivative in $d+1$ dimensions, with $A_\mu=(A_\tau,\b{A})$ the $U(1)$ gauge field and $\partial_\mu=(\partial_\tau,\nabla)$. We define
\begin{align}\label{da}
d_a(-i\b{D}) =& -\sqrt{\frac{d}{2(d-1)}}
\Lambda_{a}^{ij}D_iD_j,
\end{align}
where the $\Lambda_a$ are the $N_\Gamma$ real, symmetric, traceless $d\times d$ Gell-Mann matrices, introduced in $d=3$ to describe the band structure of $j=3/2$ Luttinger semimetals~\cite{murakami04} and later generalized to arbitrary $d$~\cite{janssen15}.

The Lagrangian $\c{L}_A$ of the gauge sector contains the usual Maxwell term, but also a derivative coupling for the magnetic field $\b{B}=\nabla\times\b{A}$:
\begin{align}\label{LA}
\c{L}_A&=\frac{1}{2}\b{E}^2+\frac{c^2}{2}\b{B}^2+\frac{v^2}{2}(\nabla\times\b{B})^2,
\end{align}
where $\b{E}=\partial_\tau\b{A}-\nabla A_\tau$ denotes the electric field. Equivalently, introducing the electromagnetic field strength tensor $F_{\mu\nu}=\partial_\mu A_\nu-\partial_\nu A_\mu$, we obtain:
\begin{align}
\c{L}_A=\frac{1}{2}F_{\tau i}^2+\frac{c^2}{4}F_{ij}^2+\frac{v^2}{4}F_{ij}(-\nabla^2)F_{ij},
\end{align}
which is suitable for generalization to $d$ spatial dimensions. To understand why the derivative term is necessary, we consider Lifshitz scaling $[\tau]=-z$, $[x_i]=-1$ with dynamic critical exponent $z$. The tree-level scaling dimensions of the
various fields and couplings of the theory follow from power counting:
\beq
&\left[\psi_\alpha\right] = d/2,  
\ \left[A_\tau\right] = (d+z-2)/2, 
\ \left[A_i\right] = (d-z)/2, \lb
&\left[e^2\right] = 2+z-d,
\ \left[m^{-1}\right] = z-2,
\ \left[c^2\right] = 2z-2,\lb
& \left[v^2\right] = 2z-4. 
\eeq
In the spatial dimension of interest $d=3$ with $z=2$, the theory is strongly coupled with relevant gauge coupling $[e^2]=1$ as in the original LAB analysis~\cite{moon13,abrikosov1971,abrikosov1974}. The theory becomes perturbatively renormalizable in $d=4-\epsilon$ with $\epsilon$ treated as a small parameter. However, the photon stiffness $v^2$ is also marginal for $z=2$ and must be included in the analysis; this coupling did not feature in the static theory without transverse gauge fields. Likewise, the photon velocity squared $c^2$ becomes a relevant coupling. The photon couplings $c^2$ and $v^2$ can be treated non-perturbatively at the level of the propagator (see App.~\ref{app:photon}).

In principle, the theory admits another relevant coupling in $d<4$: the Zeeman coupling,
\begin{align}\label{Lint}
\c{L}'=-\frac{g}{2}\sum_{\alpha=1}^{N_f}\psi_\alpha^\dagger\Sigma^{ij}\psi_\alpha F_{ij},
\end{align}
where
\begin{align}
\Sigma^{ij}=-\Sigma^{ji}=\frac{1}{4}[\Lambda_a,\Lambda_b]^{ij}\Gamma^{ab},
\end{align}
is a generator of $SO(d)$ spatial rotations in the $d_\Gamma$-dimensional spinor representation (see App.~\ref{app:Zeeman}). Here we define the $SO(N_\Gamma)$ generators $\Gamma^{ab}=\frac{1}{2i}[\Gamma^a,\Gamma^b]$. Equation~(\ref{Lint}) can be viewed as the spatial part of the Lorentz-invariant Pauli coupling in non-minimal extensions of QED~\cite{djukanovic2018,gies2020}. In $d=3$, one can check using explicit representations of the $\Lambda^a$ and $\Gamma^a$ matrices~\cite{murakami04,janssen15} that $\Sigma^{ij}=\epsilon_{ijk}J_k$ where $J_x,J_y,J_z$ are spin-3/2 matrices; thus Eq.~(\ref{Lint}) reduces to the usual Zeeman term $-g\psi^\dag\b{J}\psi\cdot\b{B}$. Note that $\b{B}$ is not a background magnetic field but a fluctuating field; $\mathcal{L}'$ is thus an interaction term, which additionally preserves time-reversal symmetry. The Zeeman coupling has dimension $[g]= (3z-d-2)/2=\epsilon/2$, which is relevant for $d<4$.

However, the Zeeman coupling can be excluded if we impose invariance of the Lagrangian under the following particle-hole transformation~\cite{boettcher16}:
\begin{align}\label{PHS}
\psi_\alpha\rightarrow(\psi_\alpha^\dag)^T,\hspace{5mm}
A_\mu\rightarrow-A_\mu,\hspace{5mm}
\Gamma^a\rightarrow-\Gamma^{a*}.
\end{align}
One can verify that under the substitution (\ref{PHS}), the fermion Lagrangian (\ref{Lpsi}) remains invariant (see App.~\ref{app:PHS}). The photon Lagrangian (\ref{LA}) is quadratic in $A_\mu$ and thus also obviously invariant. However, the Zeeman term (\ref{Lint}) is odd under this transformation. The transformation also prohibits a quadratic kinetic term $\propto\psi_\alpha^\dag(-\b{D}^2)\psi_\alpha$ proportional to the identity in spin space. The transformation (\ref{PHS}) is not a standard symmetry operation, because it involves the replacement $\Gamma^a\rightarrow\tilde{\Gamma}^a$ where $\tilde{\Gamma}^a=-\Gamma^{a*}$ forms an inequivalent representation of the Clifford algebra. Such a representation cannot be obtained via conjugation, i.e., there is no unitary charge conjugation matrix $C$ such that $C\Gamma^aC^{-1}=-\Gamma^{a*}$ holds for all $a=1,\ldots,5$. However, we have verified that at one-loop order, neither the spin-independent kinetic term nor the Zeeman term are generated if they are absent in the bare Lagrangian. (Conversely, the spin-independent kinetic term is generated to quadratic order in the Zeeman coupling.) At least at the one-loop level, the set of couplings $e^2$, $m^{-1}$, $c^2$, and $v^2$ is thus closed under RG.

\section{RG analysis}
\label{rg}

To elucidate the infrared fate of the theory, we employ the standard Wilsonian or momentum-shell RG scheme. We integrate out the high-energy modes of the action $S=\int d\tau d^dx\,\c{L}$ near the upper-critical dimension $d=4$. Due to the spacetime anisotropy of the theory, we integrate over all frequencies $-\infty<\omega<\infty$ while spatial momenta are integrated only over an infinitesimal shell $\Lambda/b<|\b{p}|<\Lambda$~\cite{herbut2007}. Here $\Lambda$ is an ultraviolet (UV) cutoff and $b=1+d\ell$ is a scale parameter where $d\ell$ is a positive infinitesimal. The effective Lagrangian after mode elimination is:
\begin{align}\label{Leff}
\mathcal{L}^<&=\sum_{\alpha=1}^{N_f}\psi_\alpha^{<\dagger}\left(Z_1D_\tau+Z_2\frac{\Gamma^a}{2m}d_a(-i\b{D})\right)\psi_\alpha^<\nn\\
&\phantom{=}+\frac{Z_3}{2}(F_{\tau i}^<)^2+\frac{Z_4c^2\Lambda^2}{4}(F_{ij}^<)^2+\frac{Z_5v^2}{4}F_{ij}^<(-\nabla^2)F_{ij}^<,
\end{align}
where the $Z_i$ are renormalization constants, related to anomalous dimensions $\gamma_i$ via $Z_i-1=\gamma_i\ln b\approx\gamma_i d\ell$, and the slow fields with momenta $|\b{p}|<\Lambda/b$ are labeled with the superscript $<$. Gauge invariance implies that the gauge charge $e$ does not undergo renormalizations independent of $Z_1$ and $Z_2$ (Ward identity). We have also introduced a factor $\Lambda^2$ in front of the $\b{B}^2$ term to redefine the relevant coupling $c^2$ as dimensionless.

\begin{figure}[t]
  \includegraphics[width=\columnwidth]{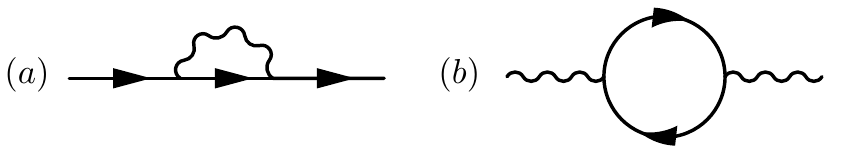}
   \caption{One-loop 1PI diagrams contributing to (a) the fermion self-energy, (b) the photon self-energy.}
   \label{pert_theory}
\end{figure}

The anomalous dimensions $\gamma_i$ are computed diagrammatically at one-loop order (Fig.~\ref{pert_theory}). Only one-particle irreducible (1PI) diagrams contribute. Fermion propagators are denoted by straight lines, and photon propagators by wavy lines. The fermion-photon QED vertex is proportional to the gauge charge $e$. The quadratic derivative coupling in Eq.~(\ref{da}) involves a quartic ``seagull'' vertex proportional to $e^2$ (see Fig.~\ref{fig:feynman2} in App.~\ref{rg_full}); due to a Ward identity this vertex does not contribute independent renormalizations and is not drawn explicitly. The $\gamma_1,\gamma_2$ anomalous dimensions arise from fermion self-energy corrections [Fig.~\ref{pert_theory}(a)], and $\gamma_3,\gamma_4,\gamma_5$, from photon self-energy corrections [Fig.~\ref{pert_theory}(b)]. As mentioned above, renormalizations of the QED vertex follow from gauge invariance and need not be computed separately. Details of the diagrammatic computations are given in App.~\ref{rg_full}.

In the second step of the Wilsonian renormalization procedure, the space and
time coordinates are rescaled according to $x'_i= b^{-1}x_i$ and
$\tau'=b^{-z}\tau$, respectively, to restore the original UV
cutoff $\Lambda$. Finally, the effective Lagrangian is brought to its original form by rescaling the fields,
\begin{align}
\psi_\alpha'&=b^{d/2}Z_1^{1/2}\psi_\alpha^<,\nn\\
A_\tau'&=b^{(d+z-2)/2}Z_3^{1/2}A_\tau^<,\nn\\
A_i'&=b^{(d-z)/2}Z_3^{1/2}A_i^<,
\end{align}
and the couplings $e^2,m^{-1},c^2$, and $v^2$. Taking derivatives of the renormalized couplings with respect to the RG scale, we obtain RG flow equations for those couplings. The RG equation for the inverse fermion band mass $m^{-1}$ is:
\begin{align}
\frac{dm^{-1}}{d\ell}=(z-2-\gamma_1+\gamma_2)m^{-1}.
\end{align}
By introducing a modified dynamic critical exponent:
\beq\label{Modifiedz}
z=2+\gamma_1-\gamma_2,
\eeq
the flow of $m^{-1}$ can be arrested. By appropriate
rescaling of the fields and the couplings, the band
parameter $m^{-1}$ 
can be scaled away from the action altogether.

Following the 
rescaling, we introduce a set of redefined parameters:
the charge coupling squared $\alpha=me^2/(8\pi^2)$, the squared
photon velocity $K=m^2c^2$, and the photon stiffness $\rho=m^2 v^2$.
The respective tree-level scaling dimensions of these redefined couplings
are $[\alpha]=4-d=\epsilon$,
$[K]=2$, and $[\rho]=0$. As mentioned before, the QED interaction can be treated perturbatively in a controlled manner in the $\epsilon$ expansion, while the photon velocity and stiffness terms are treated non-perturbatively at the propagator level. The RG flow equations for these three couplings read: 
\beq 
\frac{d\alpha}{d\ell}&=\left(\epsilon+\gamma_1-\gamma_2-\gamma_3\right)\alpha,\lb 
\frac{d K}{d\ell}&=
\left(2+2\gamma_1-2\gamma_2-\gamma_3+\gamma_4\right)K,\lb
\frac{d\rho}{d\ell}&=\left(2\gamma_1-2\gamma_2
-\gamma_3+\gamma_5\right)\rho.
\label{rge}
\eeq 
The full expressions of the anomalous dimensions $\gamma_i$
are cumbersome
and are provided in App.~\ref{rg_full}. They are
perturbative in the charge coupling but non-perturbative in the photon velocity and stiffness. To maintain perturbative control, we look for RG fixed points, i.e., common zeros of the four equations (\ref{rge}), with $\alpha_*\sim\c{O}(\epsilon)$.

Our strategy for identifying fixed points rests on the observation that the flow equations (\ref{rge}) are similar in structure to the well-known Wilson-Fisher RG equations of $\phi^4$ theory~\cite{QPT}. The photon velocity squared $K$ is a relevant coupling that plays a role analogous to the scalar mass squared $r$ in that theory. Its tree-level dimension is shifted at one-loop order by perturbative corrections of order $\epsilon$; for sufficiently large initial values of $K$, the RG equations undoubtedly admit trajectories where $K$ flows to $+\infty$ or $-\infty$ in the infrared limit $\ell\rightarrow\infty$. In Wilson-Fisher theory, $r\rightarrow-\infty$ corresponds to the broken (ferromagnetic) phase while $r\rightarrow+\infty$ corresponds to the symmetric (paramagnetic) phase of $\phi^4$ theory. While a fixed point with $K\rightarrow-\infty$ here is also identified as a broken-symmetry phase, as will be argued in Sec.~\ref{SSB}, a fixed point with $K\rightarrow+\infty$ corresponds to instantaneous photon propagation and reduces in fact to the (symmetric) LAB fixed point, as discussed in detail in the next section (Sec.~\ref{lab}).

In $\phi^4$ theory, an unstable fixed point with $r_*\sim\c{O}(\epsilon)$ separates the two phases with $r\rightarrow\pm\infty$; it is the celebrated Wilson-Fisher fixed point, which corresponds to a continuous phase transition between the two phases. This fixed point exists in the $\epsilon$ expansion because the one-loop correction term in the flow equation for $r$ is linear in the $\phi^4$ interaction strength $u$~\cite{QPT}:
\begin{align}\label{WF}
\frac{dr}{d\ell}=2r+\left(\frac{n+2}{6}\right)u+\ldots,
\end{align}
for an $n$-component field, where $\ldots$ denotes higher-order, $\c{O}(\epsilon^2)$ terms. Since $u_*\sim\c{O}(\epsilon)$ at the Wilson-Fisher fixed point, a solution $r_*=-(n+2)u_*/12\sim\c{O}(\epsilon)$ is possible. Diagrammatically, the term proportional to $u$ on the right-hand side of Eq.~(\ref{WF}) arises because a scalar mass can be generated from a $\phi^4$ interaction via a tadpole diagram. Likewise here, as will be seen in Sec.~\ref{mcp}, a photon velocity term can be generated perturbatively from the QED interaction via a polarization bubble. For small $K$, the RG equation for $K$ has the same structure as Eq.~(\ref{WF}), and an unstable fixed point with $K_*\sim\c{O}(\epsilon)$ becomes possible. This fixed point intervenes between the $K\rightarrow\pm\infty$ fixed points and, similarly to the Wilson-Fisher fixed point of $\phi^4$ theory, corresponds to a (multi)critical point of the theory, associated with a symmetry-breaking transition out of the LAB phase.

\section{LAB phase and its stability}
\label{lab}

\begin{figure}[t]
    \includegraphics[width=0.49\columnwidth]{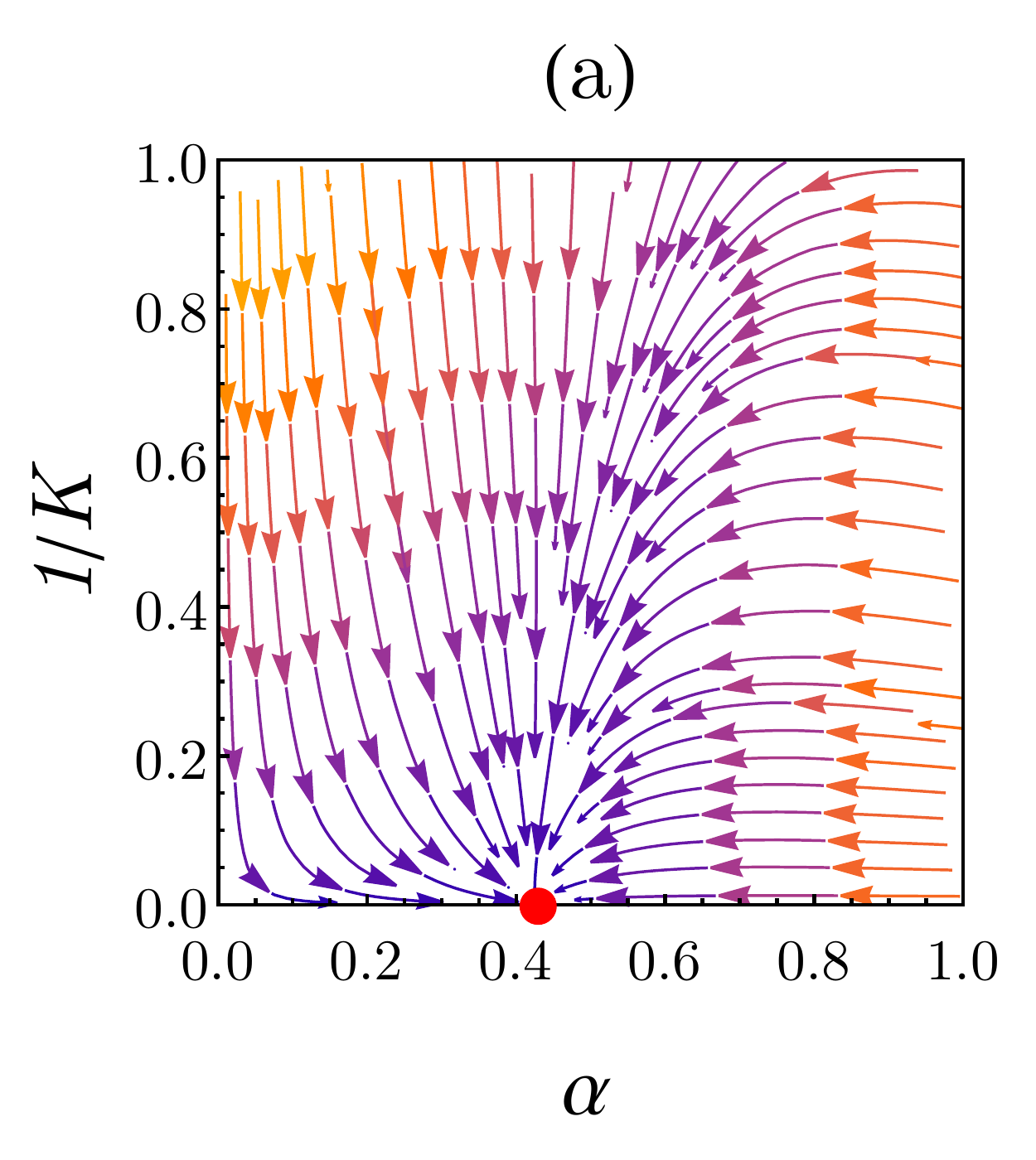}
  \includegraphics[width=0.49\columnwidth]{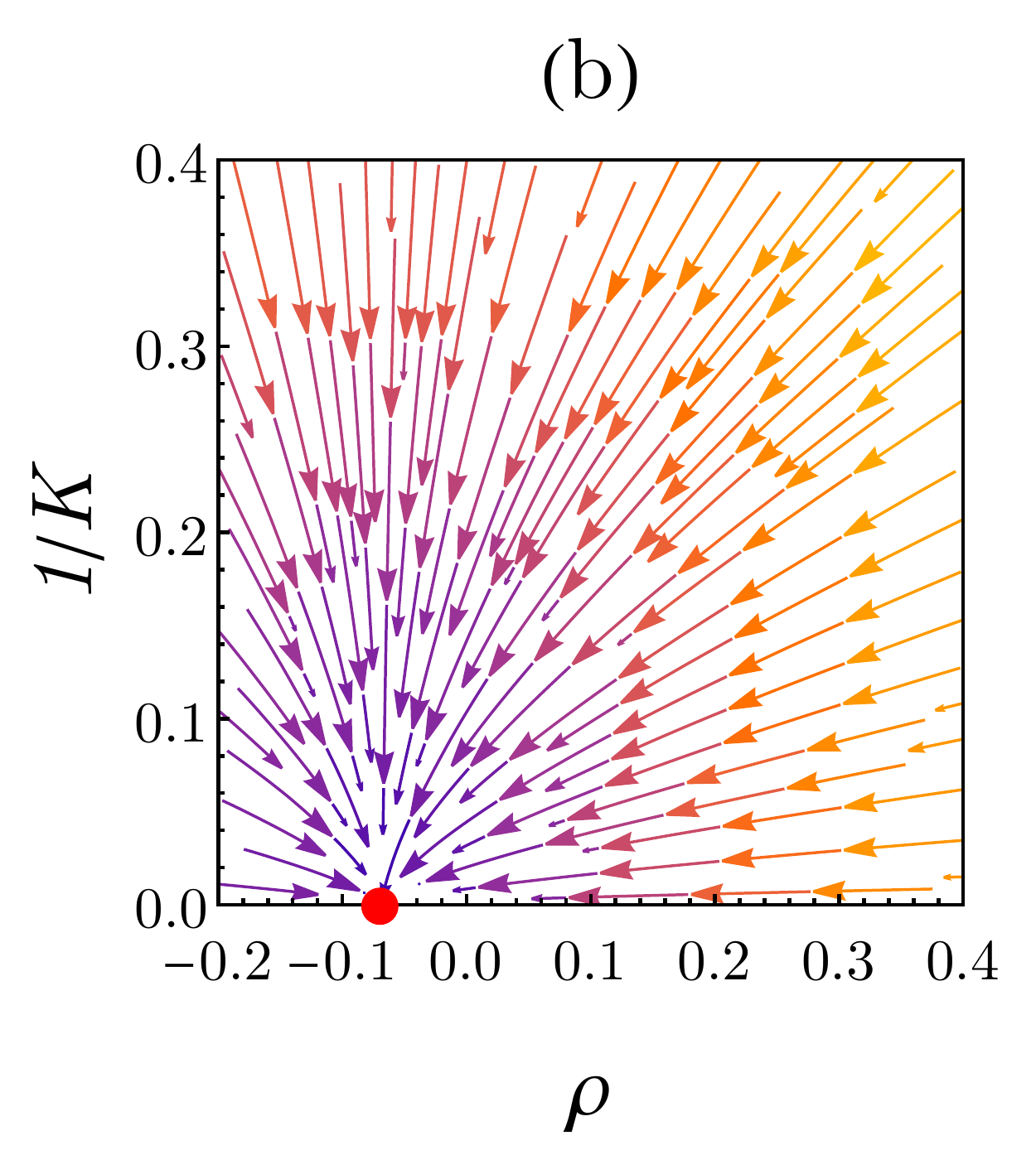}\\
  \includegraphics[width=0.49\columnwidth]{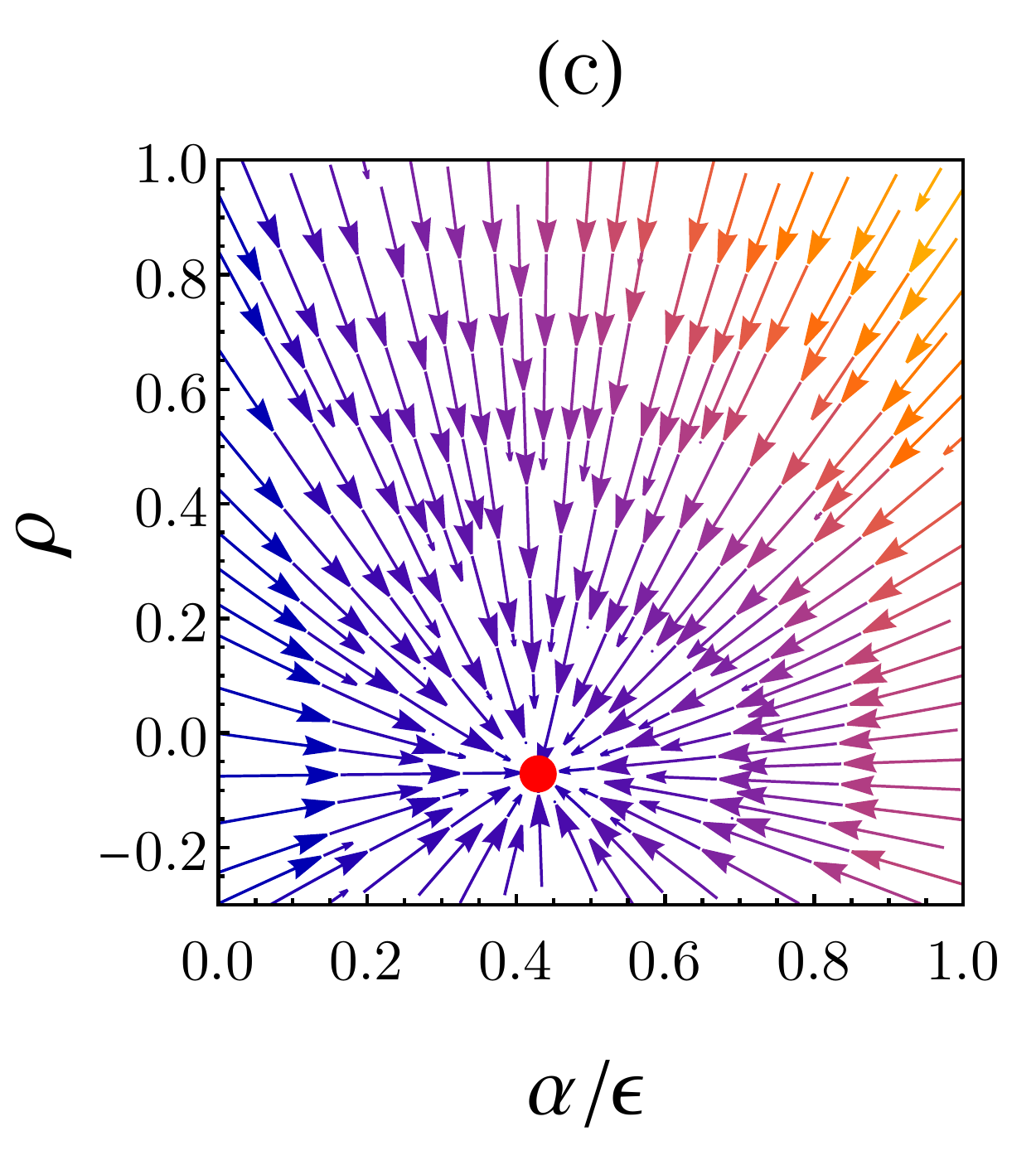}
\\
   \caption{RG flows of the gauged Luttinger fermion theory with $N_f=1$ near the LAB fixed point (red dot) in (a) the $\alpha$-$K^{-1}$ plane, (b) the $\rho$-$K^{-1}$ plane, and (c) the $\alpha$-$\rho$ plane. For (a) and (b), we set $\epsilon=1$. Flows are qualitatively similar for other values of $N_f>1$.}
   \label{crit_flow}
\end{figure}

Based on the above reasoning, we first investigate the possibility of a fixed point with infinite photon velocity. Taking the limit $K\rightarrow+\infty$ in Eq.~(\ref{rge}), we arrive at the following reduced flow equations:
\beq\label{ReducedRGEs}
\frac{d\alpha}{d\ell}
&=\left(\epsilon-\frac{6N_f+1}{3}\alpha\right)\alpha,\lb
\frac{d\rho}{d\ell}
&=-\frac{\left(5N_f+18\left(3N_f+1\right)\rho\right)}{27}\alpha.
\eeq
Besides the noninteracting fixed point $\alpha_*=0$, which is unstable, the only other fixed point of these RG equations is given by: 
\beq\label{LABFP}
&\alpha_*=\frac{3\epsilon}{6N_f+1}, \
\rho_*=-\frac{5N_f}{54N_f+18},
\eeq
with the inverse photon velocity squared $1/K_*=0$. This fixed point corresponds to the LAB fixed point~\cite{moon13,abrikosov1971,abrikosov1974,LABnote}. Although $\rho_*<0$, since $K_*=\infty$, the Lishitz term can be neglected in the photon Lagrangian (\ref{LA}), and the fixed-point action is physical. Evaluating Eq.~(\ref{Modifiedz}) at the fixed point, we find it exhibits Lifshitz scaling with dynamic critical exponent $z=2-\epsilon/(6N_f+1)$~\cite{LABnote}.

The stability of the fixed point is
given by the Hessian matrix $M_{ij}=\partial\beta_i/\partial \lambda_j $,
where $\lambda_i\in\{\alpha,1/K,\rho\}$ are the three independent
couplings. At the LAB fixed point (\ref{LABFP}), the matrix
is given by:
  \beq \label{HessianLAB}
  M&= 
  \begin{pmatrix} 
-\epsilon   & 0  & 0   \\
0 & -2+\left(\frac{6N_f+2}{6N_f+1}\right)\epsilon  & 0   \\
0 & \frac{5N_f\sqrt{K_*}\epsilon }{32(18 N_f^2+9 N_f+1)} & -\left(\frac{6N_f+2}{6N_f+1}\right)\epsilon 
\end{pmatrix},
   \eeq
to $\c{O}(\epsilon)$, with its eigenvalues:
\begin{align}\label{EigenvaluesLAB}
&-\epsilon,\hspace{5mm}-2+\left(\textstyle\frac{6N_f+2}{6N_f+1}\right)\epsilon,\hspace{5mm}
-\left(\textstyle\frac{6N_f+2}{6N_f+1}\right)\epsilon,
\end{align}
which are all negative, demonstrating stability of the fixed point for all $N_f$. Although the stability matrix (\ref{HessianLAB}) contains a diverging entry in the fixed-point limit $K_*\gg 1$, this does not affect the eigenvalues (\ref{EigenvaluesLAB}). We have also verified by numerical integration of the full RG equations (\ref{rge}) that the flow in all directions is attractive towards the LAB fixed point in its vicinity (Fig.~\ref{crit_flow}).

Stability of the LAB fixed point with respect to the photon velocity direction with a strongly irrelevant eigenvalue $-2+\c{O}(\epsilon)$ is expected at such a strong-coupling fixed point. However, irrelevance with respect to the other couplings is not {\it a priori} obvious but results from a delicate balance between tree-level dimensions and quantum corrections. Stability with respect to the charge coupling follows from stability of the LAB non-Fermi liquid in the static Coulomb limit~\cite{moon13,abrikosov1971,abrikosov1974}, but stability with respect to the $\rho$ direction, which is marginal at tree level, follows from the particular sign of the one-loop contribution to $d\rho/d\ell$ in Eq.~(\ref{ReducedRGEs}).

Our computation shows that the LAB fixed point is not only a stable phase of electrons interacting via the static Coulomb interaction, but indeed a stable phase of a full-fledged $U(1)$ gauge theory of Luttinger fermions and propagating transverse photons. In the rotationally invariant approximation, this fixed point can be understood as a stable (3+1)-dimensional critical spin liquid with QBC spinons.

\section{Broken-symmetry phase}
\label{SSB}

We now discuss the other strong-coupling fixed point with $K\rightarrow-\infty$. In $\phi^4$ theory, the $r\rightarrow-\infty$ limit is interpreted as a ferromagnetic phase. Indeed, for $r<0$ the global minimum of the interaction potential $V(\phi)=-\frac{|r|}{2}\phi^2+\frac{u}{4!}\phi^4$, with $u>0$, is at $\phi_0=\sqrt{6|r|/u}$. The symmetric vacuum $\langle\phi\rangle=0$ becomes unstable, and the ground state develops a symmetry-breaking expectation value $\langle\phi\rangle=\phi_0\neq 0$. Likewise here, we can elucidate the physical meaning of the $K\rightarrow-\infty$ fixed point by analyzing the photon Lagrangian (\ref{LA}) for $c^2=K/m^2<0$. When $c^2<0$, a paramagnetic ground state with $\langle\b{B}\rangle=0$ becomes unstable: the photon velocity becomes imaginary, which signals that one is expanding about the wrong vacuum (unstable maximum instead of stable minimum). This is analogous to how imaginary phonon frequencies ($\omega_\b{q}^2<0$) in a first-principles calculation signal a dynamical instability of the crystal structure. To determine the correct vacuum in this regime, we search for equilibrium ground states such that $\langle\b{E}\rangle=0$. To stabilize the free energy, we must add higher-order (dangerously irrelevant) terms in the Landau expansion,
\begin{align}\label{mexican}
\c{L}_A=-\frac{|c^2|}{2}\b{B}^2+\frac{v^2}{2}(\nabla\times\b{B})^2+\frac{w}{4!}(\b{B}^2)^2+\ldots,
\end{align}
with $w>0$. Restricting ourselves to uniform ground states, the second term and other possible terms with spatial derivatives can be ignored. The free energy is minimized by the fluctuating magnetic field acquiring a nonzero expectation value $\langle\b{B}\rangle=B_0\hat{\b{e}}$ in some direction $\hat{\b{e}}$, with magnitude $B_0=\sqrt{6|c^2|/w}$. This problem was analyzed previously in the context of quantum dimer models/$U(1)$ spin liquids in 3+1 dimensions, where this broken ``magnetic'' phase adjacent to the $c^2=0$ Rokhsar-Kivelson critical point corresponds microscopically to a valence bond crystal~\cite{moessner03,hermele04}. The Lagrangian (\ref{mexican}) is also closely related to a Lagrangian used previously to describe an isotropic-to-nematic transition in the fractional quantum Hall effect~\cite{mulligan2010}.

\begin{figure}[t]
  \includegraphics[width=0.6\columnwidth]{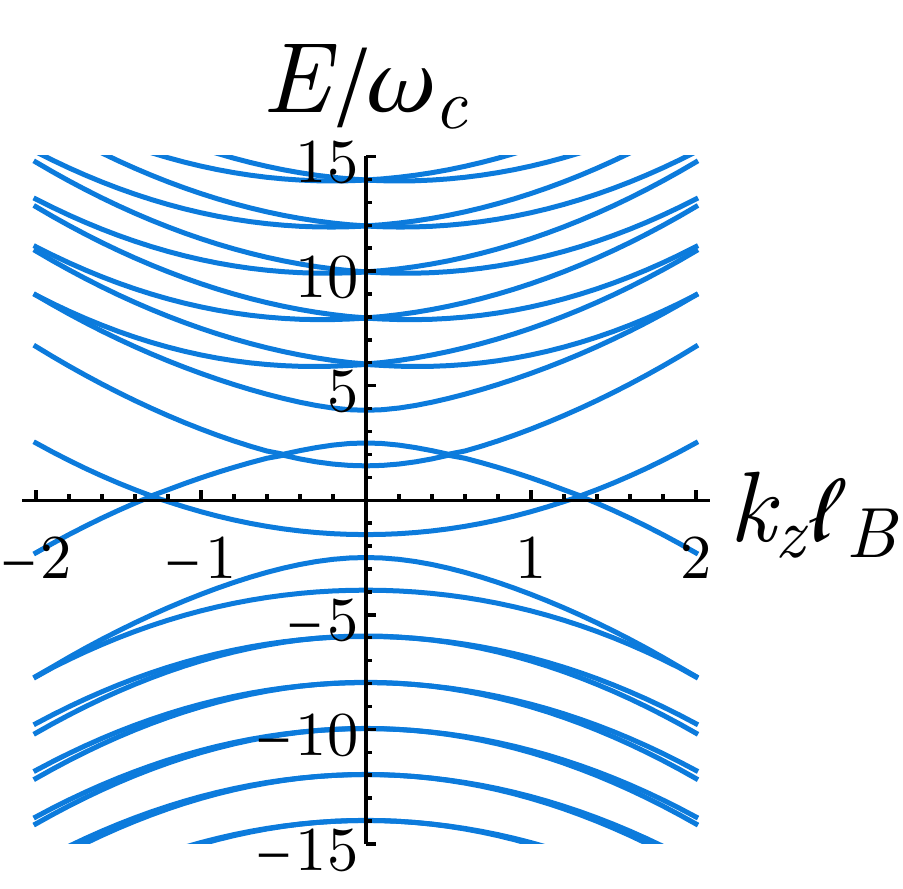}
   \caption{Landau levels in the fermion spectrum of the time-reversal symmetry-breaking phase with spontaneously generated magnetic field $\langle\b{B}\rangle=B_0\hat{\b{z}}$, where $\omega_c=eB_0/m$ is the cyclotron frequency, $k_z$ is the fermion momentum in the field direction, and $\ell_B=\sqrt{1/eB_0}$ is the magnetic length.}
   \label{fig:BLL}
\end{figure}

In the present context, the generation of a vacuum magnetic field $\langle\b{B}\rangle\neq 0$ signals the spontaneous breakdown of rotational and time-reversal symmetries in the ground state. To the difference of the quantum dimer problem, here this background magnetic field is additionally perceived by the Luttinger fermions and leads to the formation of Landau levels~\cite{rhim015}. By rotational invariance of the zero-field problem, the Landau level spectrum is independent of the field direction. In Fig.~\ref{fig:BLL}, we plot the Landau level spectrum as a function of the fermion momentum $k_z$ along the field direction ($\hat{\b{e}}=\hat{\b{z}}$), which is a good quantum number. The spectrum exhibits unusual features like level crossings, unequal level spacings, and unequal band curvatures, which have been discussed previously~\cite{rhim015}.

\begin{figure}[t]
\includegraphics[width=0.49\columnwidth]{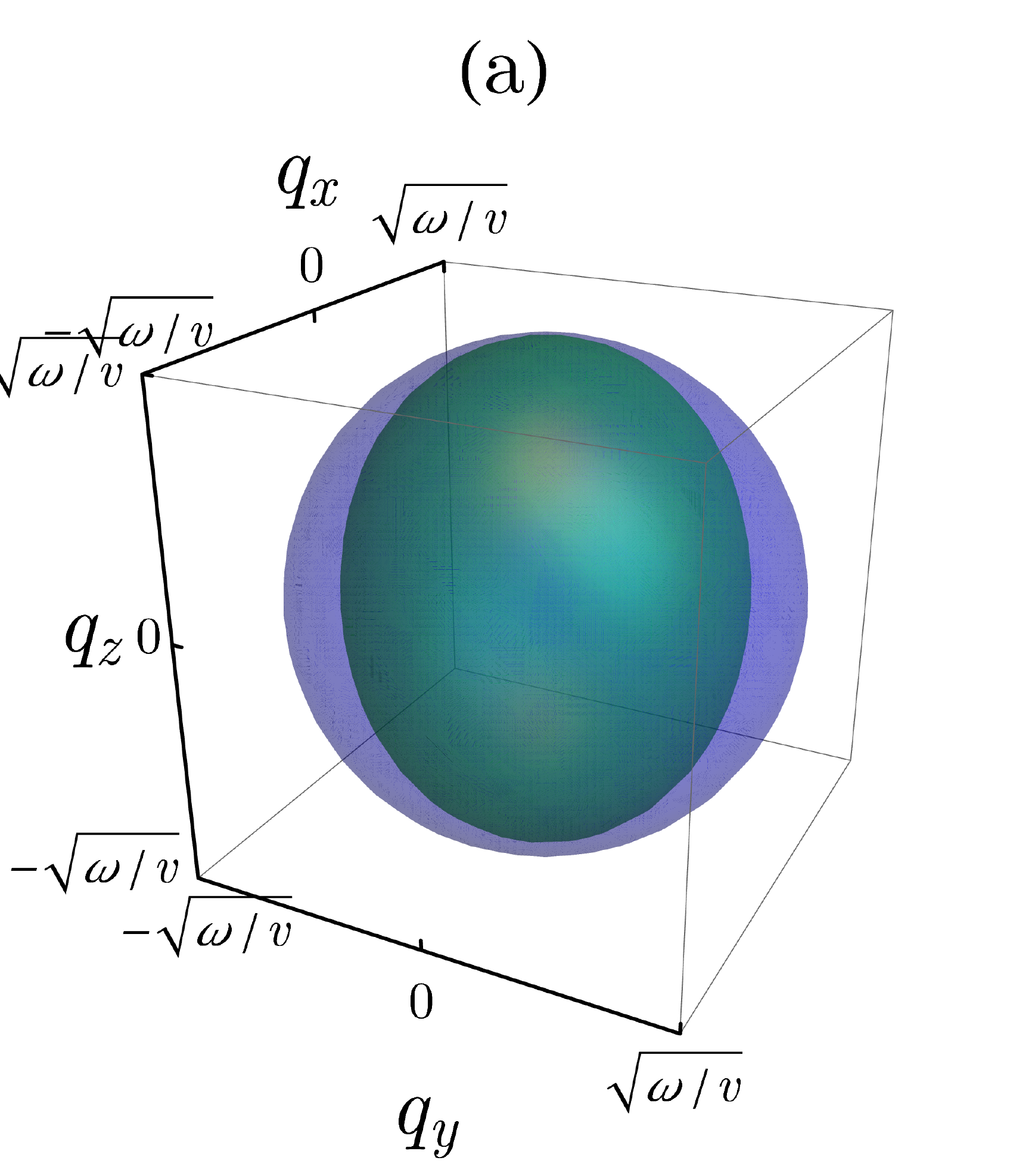}
\includegraphics[width=0.49\columnwidth]{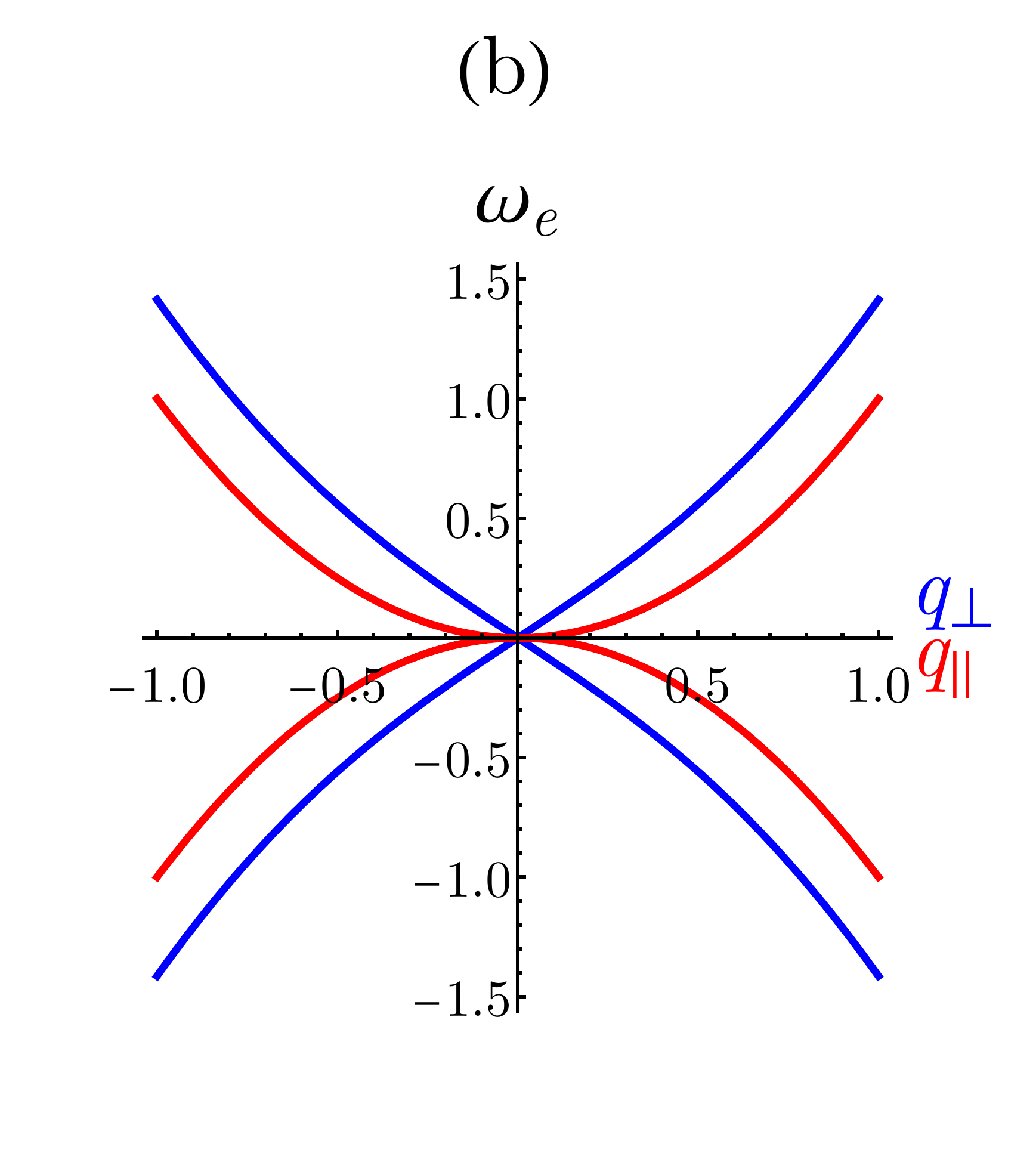}
   \caption{(a) Isofrequency surfaces of the ordinary branch (purple) and extraordinary branch (green) of the emergent photon dispersion in the broken-symmetry phase; (b) extraordinary wave dispersion in directions parallel (red) and perpendicular (blue) to the spontaneously generated magnetic field.}
   \label{fig:photon}
\end{figure}

In addition to the fermion spectrum, we may ask how the photon spectrum is affected by the background magnetic field. To derive the photon spectrum in the broken phase, we expand the electromagnetic Lagrangian (\ref{mexican}) to quadratic order in the fluctuations $\b{e}\equiv\b{E}-\langle\b{E}\rangle$, $\b{b}\equiv\b{B}-\langle\b{B}\rangle$:
\begin{align}\label{LAfluct}
\c{L}_A=\frac{1}{2}\b{e}^2+\frac{1}{2}\tilde{c}^2b_\parallel^2+\frac{v^2}{2}(\nabla\times\b{b})^2+\c{O}(b^3),
\end{align}
where we have decomposed the magnetic fluctuation about $\langle\b{B}\rangle=B_0\hat{\b{z}}$ into longitudinal and transverse parts,
\begin{align}
\b{b}=b_\parallel\hat{\b{z}}+\b{b}_\perp,
\end{align}
with $\b{b}_\perp\cdot\hat{\b{z}}=0$, and we define $\tilde{c}=\sqrt{2|c^2|}$. As expected for a broken $O(3)$ symmetry, the coefficient of the ``mass term'' $\propto\b{b}_\perp^2$ for transverse fluctuations vanishes, while it is nonzero and renormalized by a factor of two for longitudinal fluctuations. However, the mode spectrum itself remains massless by gauge invariance. Passing to momentum space $Q\equiv(\omega,\b{q})$, the quadratic action associated with Eq.~(\ref{LAfluct}) can be written in the following suggestive form:
\begin{align}
S_A=\frac{1}{2}\int\frac{d^4Q}{(2\pi)^4}\left[\delta_{ij}e_i(-Q)e_j(Q)+\mu_{ij}^{-1}(\b{q})b_i(-Q)b_j(Q)\right],
\end{align}
which describes a linear dispersive medium with isotropic permittivity tensor $\varepsilon_{ij}=\delta_{ij}$ and anisotropic permeability tensor:
\begin{align}
\mu_{ij}(\b{q})=\left(\begin{array}{ccc}
\mu_\perp(\b{q}) & & \\
& \mu_\perp(\b{q}) & \\
& & \mu_\parallel(\b{q})
\end{array}
\right),
\end{align}
with:
\begin{align}\label{permeability}
\mu_\perp(\b{q})=\frac{1}{v^2\b{q}^2},\hspace{5mm}
\mu_\parallel(\b{q})=\frac{1}{\tilde{c}^2+v^2\b{q}^2}.
\end{align}
The photon dispersion can then be extracted using classical methods~\cite{LL_ECM}. We use Maxwell's equations in momentum space and in the absence of sources,
\begin{align}
\b{q}\times\b{e}=\omega\b{b},\hspace{5mm}
\b{q}\times\b{h}=-\omega\b{d},
\end{align}
denoting the usual macroscopic fields $\b{e},\b{b},\b{d},\b{h}$ by lowercase letters. Using additionally the constitutive relations $d_i=e_i$ and $b_i=\mu_{ij}h_j$, we find:
\begin{align}
\bigl[\b{q}^2\delta_{ij}-q_iq_j-\omega^2\mu_{ij}(\b{q})\bigr]h_j(Q)=0,
\end{align}
which admits nontrivial solutions only if the determinant of the matrix on the left-hand side vanishes. Evaluating this determinant, we find that it factorizes into two possible conditions:
\begin{align}
\omega^2-\frac{\b{q}^2}{\mu_\perp(\b{q})}=0,\hspace{5mm}
\omega^2-\frac{q_x^2+q_y^2}{\mu_\parallel(\b{q})}-\frac{q_z^2}{\mu_\perp(\b{q})}=0,
\end{align}
which are the Fresnel equations for a medium with uniaxial magnetic anisotropy~\cite{LL_ECM}. Substituting (\ref{permeability}), we find two distinct modes:
\begin{align}
\omega_o(\b{q})&=\pm v\b{q}^2,\label{ordinary}\\
\omega_e(\b{q})&=\pm\sqrt{\tilde{c}^2q_\perp^2+v^2(\b{q}^2)^2},\label{extraordinary}
\end{align}
with $q_\perp^2=q_x^2+q_y^2$, and $\pm$ denotes the particle/antiparticle branches of the dispersion. In Fig.~\ref{fig:photon}(a), we display isofrequency surfaces for both modes. The broken phase thus exhibits emergent optical birefringence with an isotropic ordinary wave (\ref{ordinary}) and anisotropic extraordinary wave (\ref{extraordinary}). By contrast with conventional birefringence however, here the ordinary wave is a $z=2$ mode, while the extraordinary wave has $z=2$ scaling in the longitudinal direction but $z=1$ scaling in the transverse direction [Fig.~\ref{fig:photon}(b)]:
\begin{align}
\omega_e(0,0,q_\parallel)=\pm vq_\parallel^2,\hspace{5mm}
\omega_e(q_x,q_y,0)=\pm\tilde{c}q_\perp,
\end{align}
in the long-wavelength limit.

To summarize, the broken-symmetry phase ($r\rightarrow-\infty$ fixed point) corresponds to an unusual type of itinerant ferromagnet where gauge fluctuations lead to the spontaneous generation of a background magnetic field, which subsequently induces orbital magnetization for the fermions and birefringence of emergent photons. In the spin liquid interpretation, this can be viewed as a type of chiral spin liquid in 3+1 dimensions.

\section{Lifshitz-QED multicritical point}
\label{mcp}

Having discussed the two stable phases of the coupled fermion-photon system, the symmetric LAB phase (Sec.~\ref{lab}) and the ferromagnetic broken-symmetry phase (Sec.~\ref{SSB}), we now turn to possible unstable fixed points describing a quantum phase transition between the two. As the transition is reached by tuning the photon velocity squared $K$ through zero, we proceed as in $\phi^4$ theory~\cite{QPT} and seek a Wilson-Fisher-type unstable fixed point with $K_*\sim\c{O}(\epsilon)$, that is perturbatively accessible within the $\epsilon$ expansion. As explained at the end of Sec.~\ref{rg}, this scenario is possible if the relevant coupling receives $\c{O}(\epsilon)$ corrections from interactions at the fixed point, i.e., if the RG beta function for $K$ is similar in structure to Eq.~(\ref{WF}). Assuming $K\sim\c{O}(\epsilon)$, we find indeed that $dK/d\ell$ in Eq.~(\ref{rge}) becomes:
\begin{align}\label{dKdl}
\frac{dK}{d\ell}=2K+\frac{13N_f\alpha}{36},
\end{align}
to $\c{O}(\epsilon)$, which is precisely the same structure as Eq.~(\ref{WF}). The remaining RG equations in this limit are:
\begin{align}
\frac{d\alpha}{d\ell}&=\epsilon\alpha-[2N_f+f_1(\rho)]\alpha^2,\label{dadl}\\
\frac{d\rho}{d\ell}&=f_2(\rho,N_f)\alpha,\label{drhodl}
\end{align}
to leading order in $\epsilon$. The functions $f_1$ and $f_2$ are given in App.~\ref{fixeds}. From Eq.~(\ref{dKdl}), the Wilson-Fisher scenario is realized by the fixed-point value:
\begin{align}
K_*=-\frac{13N_f\alpha_*}{72},
\end{align}
which is $\c{O}(\epsilon)$ if the fixed-point interaction $\alpha_*$ is. Note that, like in $\phi^4$ theory where $r_*<0$, here also $K_*<0$, expressing the fact that the position of the critical point is shifted from its bare value ($K=0$) by one-loop corrections.

\begin{figure}[t]
  \includegraphics[width=0.6\columnwidth]{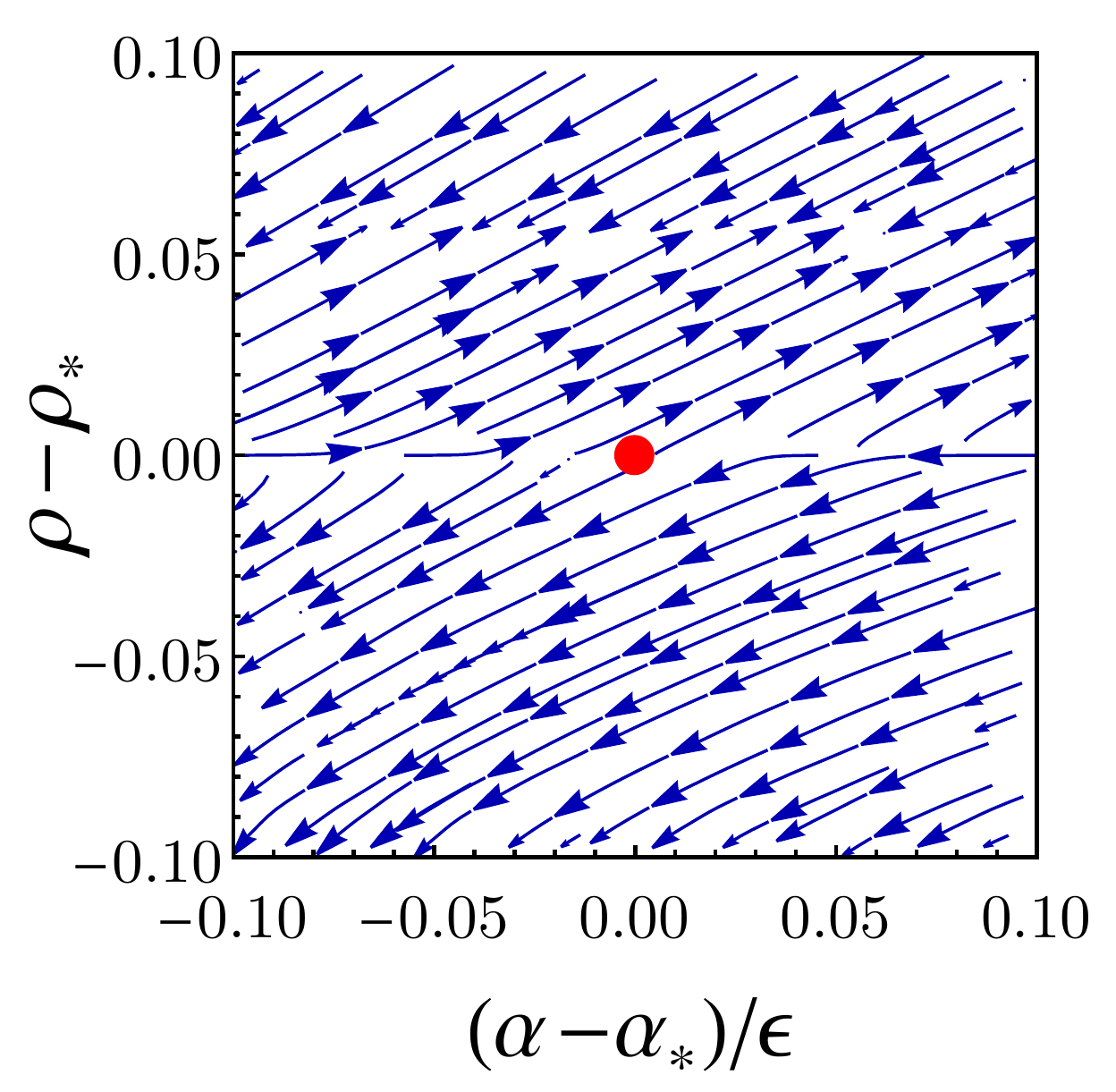}
   \caption{RG flows of the gauged Luttinger fermion theory in the vicinity of the Lifshitz-QED (LQED) multicritical point $\alpha_*\neq 0$, $\rho_*\neq 0$ (red dot) for the threshold fermion flavor number $N_f=2$, in the $\alpha$-$\rho$ plane. The fixed point is stable along the $\alpha$ direction but unstable along the remaining directions. RG flows are qualitatively similar for other values of $N_f\geq 2$.}
   \label{lqed_flow}
\end{figure}

To obtain a physical fixed point, we search for common zeros $(\alpha_*,\rho_*)$ of the beta functions (\ref{dadl}-\ref{drhodl}) with both fixed-point couplings positive. Furthermore, $\alpha_*$ is $\c{O}(\epsilon)$ while $\rho_*\sim\c{O}(1)$. For $N_f=1$, we do not find any such fixed points except the trivial Gaussian fixed point $\alpha_*=K_*=0$ and $\rho_*$ arbitrary, which exists for arbitrary $N_f$. This describes a decoupled system of free Luttinger fermions and free Lifshitz photons with $z=2$, which is RG-unstable in all directions.

\begin{figure}
  \includegraphics[width=0.99\columnwidth]{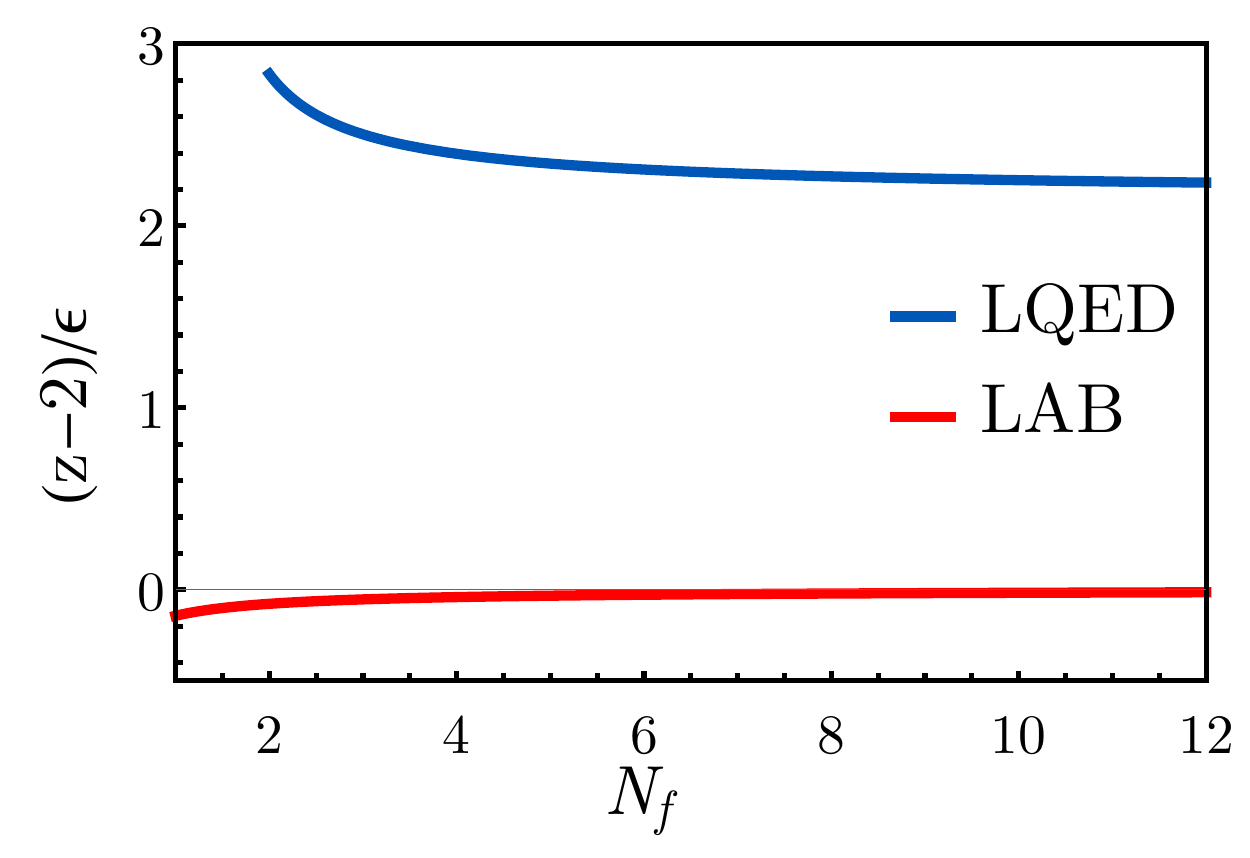}
   \caption{Departure of the dynamic critical exponent $z$ from its Gaussian-limit value of $z=2$ for the LAB (all $N_f$) and LQED ($N_f\geq 2$) fixed points, as a function of number of fermion flavors $N_f$.}
   \label{dyn_exp}
\end{figure}

For $N_f\geq 2$, we find an unstable fixed point with finite photon stiffness $\rho_*>0$ and finite gauge coupling $\alpha_*>0$:
\beq
\alpha_*=\frac{\epsilon}{2N_f+f_1(\rho_*)}, \ \
f_2(\rho_*,N_f)=0,
\eeq
where the value of $\rho_*$ is given by the solution of the second equation. That equation is highly nonlinear, since $\rho$ is not a perturbative coupling, and must be solved numerically for each $N_f$ (see App.~\ref{fixeds}). Close to this Lifshitz-QED (LQED) fixed point, the RG flow is attractive along the $\alpha$ direction but repulsive along the $\rho$ direction (Fig.~\ref{lqed_flow}), as well as the $K$ direction. In App.~\ref{fixeds}, we further investigate the stability of the LQED fixed point by computing its stability matrix. This multicritical point controls a special critical line on a two-dimensional first-order phase boundary between the LAB phase and the broken-symmetry phase where the transition becomes continuous (Fig.~\ref{phase_diag}). The emergent Lifshitz quantum electrodynamics
at the fixed point is described by the following critical action:
\begin{align}\label{lqed}
\c{L}_\text{LQED}^*&=\sum_{\alpha=1}^{N_f}\psi_\alpha^\dagger\left(D_\tau+\frac{\Gamma^a}{2m}d_a(-i\b{D})\right)\psi_\alpha\nn\\
&\phantom{=}+\frac{1}{2}\b{E}^2+\frac{v^2}{2}(\nabla\times\b{B})^2,
\end{align}
i.e., Luttinger fermions minimally coupled to Lifshitz photons. In the infrared, the dynamic critical exponent $z>2$ of the coupled system remains anisotropic, but deviates strongly from both its tree-level value $z=2$ and its value at the LAB fixed point (Fig.~\ref{dyn_exp}).

\section{Summary and outlook}
\label{summary}

In summary, we have investigated the infrared fate of a theory of QBC Luttinger fermions coupled to a dynamical $U(1)$ gauge field in 3+1 dimensions. Modulo the simplifying assumptions of rotational invariance and a particle-hole symmetric dispersion, RG arguments led us to augment the standard Maxwell terms in the gauge-field Lagrangian by a derivative coupling for the fluctuating magnetic field $\b{B}$. Carrying out an RG analysis near the upper-critical dimension, we discovered that the gauge theory assumes the structure of a Landau-Ginzburg-Wilson theory for a peculiar type of ferromagnetic quantum phase transition whose order parameter is the gauge-invariant magnetic field $\b{B}$.

The tuning parameter for the transition is the photon velocity squared $c^2$. The symmetric phase with uncondensed order parameter $\langle\b{B}\rangle=0$ was found to coincide with the non-Fermi liquid LAB phase studied previously. Its stability with respect to transverse gauge fluctuations, elucidated here, indicates it can be viewed as a novel critical spin liquid phase in 3+1 dimensions, analogous in some respects to the algebraic spin liquid in 2+1 dimensions. In particular, such a QBC spin liquid in 3+1 dimensions has the advantage over its (2+1)-dimensional counterpart~\cite{xu2012} that monopole condensation is a strong-coupling phenomenon in 3+1 dimensions, while instanton proliferation is a weak-coupling phenomenon in 2+1 dimensions~\cite{polyakov1987}. Extending the present work to compact $U(1)$ gauge fields, a QBC spin liquid is thus expected to be inherently more stable in 3+1 dimensions than in 2+1 dimensions.

Turning to the broken phase with $\langle\b{B}\rangle\neq 0$, it corresponds to a time-reversal symmetry-breaking spin liquid with spinon Landau levels and birefringent emergent photons. Adapting the Wilson-Fisher paradigm of $\phi^4$ theory to the current context, we found for $N_f\geq 2$ a multicritical point intervening between the two phases at a critical value $c_*^2\sim\c{O}(\epsilon)$ of the photon velocity squared. This multicritical point, with $z\neq 2$ anisotropic scaling for both fermions and photons, represents a novel example of strongly-coupled Lifshitz gauge theory~\cite{nishida07,kachru2008}.

Several questions remain open to further research. Perturbations that break particle-hole symmetry or continuous rotational invariance in the QBC dispersion~\cite{moon13,boettcher16,boettcher17} significantly enlarge the dimension of the coupling-constant space and may further destabilize the fixed points studied here. The detailed phenomenology of the broken phase should be further investigated, in particular, the effect of interactions between low-energy Landau-level fermions and anisotropic photons. It would also be desirable to complement the continuum field-theoretic analysis presented here with microscopic lattice model realizations, for example starting with a microscopic spin model in three spatial dimensions and exploring QBC fermionic parton ans\"atze along the lines of Refs.~\cite{xu2012,mishmash2013}. Even at the level of continuum field theories, $1/N_f$ expansions~\cite{janssen16} and $2+\epsilon$ expansions~\cite{janssen2017b} may complement the $4-\epsilon$ expansion utilized here and shed further light on the nature of the unusual phase transition we have discussed. Additionally, if the scenario studied in Refs.~\cite{herbut2014,janssen15} remains operative in the presence of transverse gauge fluctuations, it would predict the destabilization of the LAB phase towards an interaction-induced (3+1)-dimensional strong topological insulator~\cite{zhang2009} with nematic order. Given the underlying $U(1)$ gauge structure, this would provide a natural route towards realizing the Pesin-Balents topological Mott insulator~\cite{pesin2010}, an example of fractionalized topological insulator~\cite{maciejko2015}.

\begin{acknowledgments}
We thank I. Boettcher for useful discussions. J.M. acknowledges funding from NSERC Discovery Grants \#RGPIN-2020-06999 and \#RGPAS-2020-00064; the Canada Research Chair (CRC) Program; the Government of Alberta's Major Innovation Fund (MIF); the University of Alberta; the Tri-Agency New Frontiers in Research Fund (NFRF, Exploration Stream); and the Pacific Institute for the Mathematical Sciences (PIMS) Collaborative Research Group program.
\end{acknowledgments}

\appendix

\setcounter{figure}{0}
\renewcommand{\thefigure}{A\arabic{figure}}

\onecolumngrid

\section{Action of gauged Luttinger fermions}

This section details the minimal theory of charged Luttinger
fermions in $d+1$ space-time dimensions. 
The one-loop renormalization of the interactions in
this theory leads to the results discussed in the main text.

\subsection{Fermion sector}
\label{fermion}

We consider a $U(1)$ gauge theory coupled to fermions with 
cubic-symmetry protected quadratic band crossing (QBC). In the rotation and particle-hole
symmetric limits, using $ N_\Gamma=(d+2)(d-1)/2$ 
Euclidean Clifford matrices $\{\Gamma^a\}$ 
  of dimensions $2^{\lfloor N_\Gamma/2\rfloor}$, 
  the fermionic Lagrangian of QBC fermions can be
expressed in $d$-dimensional space 
  as~\cite{murakami04,janssen15}:
\begin{align}
\L_\psi = \psi^\dagger\left[\partial_\tau
+\frac{\Gamma^a}{2m} d_a(-i\nabla)\right]\psi,\hspace{10mm}
d_a(-i\nabla) = -\sqrt{\frac{d}{2(d-1)}}
\Lambda_{a}^{ij}\partial_i\partial_j, \hspace{10mm}
\{\Gamma^a,\Gamma^b\} =2\delta^{ab},
\label{matter}
\end{align}
where $\{\Lambda_a\}$ are the $N_\Gamma$ real, symmetric 
$d\times d$ Gell-Mann matrices~\cite{janssen15}. In three-dimensional momentum space the band dispersion is explicitly 
  given by the functions,
\beq
&d_1(\b{k})=\frac{\sqrt 3}{2}\left(k_x^2-k_y^2\right), \hspace{10mm}
d_2(\b{k})=\sqrt{3}k_x k_y, \hspace{10mm}
d_3(\b{k})=\sqrt{3}k_xk_z, \nonumber\\
&d_4(\b{k})=\sqrt 3 k_y k_z, \hspace{10mm}
d_5(\b{k})=\frac{1}{2}\left(2k_z^2-k_x^2-k_y^2\right).
\eeq
In our notation, the greek indices span space-time and the roman indices span space.
Due to the Euclidean Clifford algebra, $\Gamma^{ab}=[\Gamma^a,\Gamma^b]/(2i)$ are the 
$N_\Gamma(N_\Gamma-1)/2$ generators 
of $SO(N_\Gamma)$ rotations. Certain useful identities of the $\Gamma$ 
matrices that we use are as follows:
\beq
\Gamma^b\Gamma^a\Gamma^b&=(2-N_\Gamma)\Gamma^a,\nonumber\\
\Gamma^c\Gamma^a\Gamma^b\Gamma^c&=(N_\Gamma-4)\Gamma^a\Gamma^b+4\delta_{ab},\nonumber\\
\Gamma^d\Gamma^a\Gamma^b\Gamma^c\Gamma^d&=-2\Gamma^c\Gamma^b\Gamma^a-(N_\Gamma-4)
  \Gamma^a\Gamma^b\Gamma^c.
\eeq
The symmetric Gell-Mann matrices obey the identities~\cite{janssen15},
\beq
\Lambda^{ij}_a\Lambda^{kl}_a
=\left(\delta_{ik}\delta_{jl}+\delta_{il}\delta_{jk}\right)
-\frac{2}{d}\delta_{ij}\delta_{kl},\hspace{10mm}
\tr(\Lambda_a\Lambda_b)=2\delta_{ab}.
\label{lambda_id}
\eeq
Using these identities it can be shown that
\beq
d_a(\b{p})d_a(\b{q})=
\frac{d}{d-1}\left(\left(\b{p} \cdot\b{q}\right)^2-
\frac{1}{d}
p^2q^2\right)
\implies
d_a(\b{q})d_a(\b{q})=q^4.
\eeq

The global $U(1)$ symmetry of the action Eq.~\eqref{matter} is manifest with the global transformations 
$\psi\mapsto e^{ie\alpha}\psi$. 
We elevate the global $U(1)$ symmetry of the model to a local $U(1)$ gauge symmetry
by introducing vector gauge photons $A_\mu$ that are minimally coupled with the action.
The inclusion of the minimal coupling is achieved by replacing 
all the normal derivatives of the action with covariant derivatives
$D_\mu=\partial_\mu+ieA_\mu$. With the covariant derivatives, the action reads:
\beq
\L_\psi &= \psi^\dagger\left(D_\tau-\sqrt{\frac{d}{2(d-1)}}\Lambda_a^{ij}\Gamma^a\frac{D_i D_j}{2m}\right)\psi,
\label{cov_matter}
\eeq
with the rule that under the local $U(1)$ transformations both the fermionic fields and
the vector gauge fields transform simultaneously with the formula,
\beq
\psi(X)&\mapsto e^{ie\varphi(X)}\psi(X),\nonumber\\
A_\mu(X)&\mapsto A_\mu(X) -\partial_\mu\varphi(X),
\label{gauge}
\eeq
where $X=(\tau,\b{x})$ denotes the space-time coordinate. Similarly for the frequency-momentum
space we use the notation $K=(\omega,\b{k})$. 
It is straightforward to check that under the gauge transformations, Eq.~\eqref{gauge}, the action of the covariant derivative changes as,
\beq
&D_\mu\psi(X)\mapsto (\partial_\mu+i e A_\mu(X)-ie\partial_\mu\varphi(X))
e^{ie\varphi(X)}\psi(X)
=e^{ie\varphi(X)}D_\mu\psi(X),
\eeq
which in turn ensures
that the minimally coupled action \eqref{cov_matter} 
remains invariant under the gauge transformation. The corresponding global currents are given by,
\beq
j_0&=i\psi^\dagger\psi,\nonumber\\
j_i&=i\sqrt{\frac{d}{2(d-1)}}\frac{\Lambda_a^{ij}}{2m}
\left[
  \psi^\dagger\Gamma^a D_j\psi
  -
  \left(D_j\psi\right)^\dagger\Gamma^a\psi
\right],
\label{global_current}
\eeq
with the gauge-invariant, current conservation condition,
$D_\mu j_\mu=0$ dictated by the symmetry. 

\subsection{Photon sector}
\label{app:photon}

With the fermionic sector so defined we need a gauge-invariant description of the 
gauge field action. We are interested in a $U(1)$ gauge theory in 3+1 dimensions. The gauge-invariant Euclidean Maxwell action is given by 
$\L_{\rm Maxwell}=(1/4)F^2_{\mu\nu}+e j_\mu A_\mu$,
with the matter current $j_\mu$ defined above [Eq.~\eqref{global_current}]. 
In our theory, space and time have anisotropic Lifshitz scaling in the fermionic sector.
Anticipating an anisotropic renormalization, the photon
Lagrangian is most compactly written in the Euclidean space-time as,
\beq
\L_{ U(1)} &= 
\frac{1}{2}
F_{\tau i}F_{\tau i}+
\frac{1}{4}
F_{ij}
\left(c^2-v^2\nabla^2\right)
F_{ij}
+ e j_\mu A_\mu, 
\label{gauge_act}
\eeq
with $F_{\mu\nu} = \partial_\mu A_\nu - \partial_\nu A_\mu$. Here, $c$ is the photon velocity, and 
the additional $\propto v^2$ term is kept as it is marginal from power counting in the
Lifshitz scaling limit.
As a consequence of gauge invariance, the inverse photon propagator obtained from this action is singular.
To determine the photon propagator, one needs
to work in a fixed gauge suitable to the inherent space-time asymmetry. 
We use a non-local gauge that has been proposed in a similar context~\cite{anselmi10},
which leads to a simple photon Green's function. The gauge-fixing prescription
for Maxwell electrodynamics~\cite{peskin95} can be easily adapted to
this context, as we now discuss. A generic gauge-fixing 
function for the photons is given by:
\beq
G(A_\mu,\omega)
=
\partial_\tau A_\tau(X)+L\partial_i A_i(X)-\omega(X)=0,
\label{gauge_orbit}
\eeq
where $\omega(X)$ is an arbitrary function and a differential operator 
$L=c^2-v^2\nabla^2$ is introduced.  
In the path integral
we must only sum over the unique gauge configurations dictated by the 
gauge orbit $\omega(X)$. Various different gauge field
configurations which are related to each other through 
reparametrization must be counted once. The original functional 
integral is given by:
\beq
Z_{U(1)}[j]=\int\mathcal D A_\mu\, e^{-\int_X \L_{ U(1)}(A_\mu,j_\mu)},
\eeq
where $\int_X=\int d\tau d^dx$ is a short-hand notation for 
the $d+1$ dimensional space-time integral.
We wish to integrate over gauge field configurations that satisfy the 
gauge fixing condition $G(A_\mu,\omega)=0$. Therefore, we need to 
sum over all gauge field reparametrizations 
$A_\mu \mapsto A^\alpha_\mu=A_\mu+\partial_\mu\alpha$,
that leave the functional integral invariant
but also obey the gauge fixing condition. The removal of the singularity in the 
gauge field propagator and fixing of the gauge choice can be simultaneously
carried out by the Faddeev-Popov prescription \cite{schwartz13}. In the first step, 
a coordinate-transformed delta function identity, 
\beq
1=\int\mathcal D\alpha\det\left(\frac{\delta G(A_\mu^\alpha,\omega)}{\delta\alpha}
\right)
\delta\left(G(A_\mu^\alpha,\omega)\right),
\eeq
is used 
to modify the partition function of the gauge fields by a resolution of identity,
\beq
Z_{U(1)}[j]=\int\mathcal D\alpha\int\mathcal D A_\mu\, 
e^{-\int_x \L_{ U(1)}(A_\mu,j_\mu)}
\det\left(\frac{\delta G(A_\mu^\alpha,\omega)}{\delta\alpha}
\right)
\delta\left(G(A_\mu^\alpha,\omega)\right).
\eeq
With $\delta G(A_\mu^\alpha(X),\omega)/\delta\alpha(Y)
=\left(\partial_\tau^2+L\nabla^2\right)\delta^{(d+1)}(X-Y)
$, the determinant in this expression factors out. The non-local gauge 
choice adopted in Ref.~\cite{anselmi10} is carried out 
with a gauge fixing function,
\beq
G(A_\mu,\varphi)
=\partial_\tau A_\tau+L\partial_i A_i
-L^{1/2}\varphi,
\eeq
where we sum over all functions $\varphi(X)$ with a Gaussian weight centered around
$\varphi(X)=0$. Finally, we make a gauge transformation 
$A_\mu\mapsto A_\mu+\partial_\mu\alpha$ in the functional integral over the gauge fields 
to obtain,
\begin{align}
Z_{ U(1)}[j]
&=\int\mathcal D\varphi\, e^{-
\int_X\frac{\varphi(X)^2}{2\xi}}
\int\mathcal D\alpha
\det\left(\partial_\tau^2+L\nabla^2\right)
\int \mathcal D A^\alpha_\mu\,
e^{-\int_x\mathcal L_{U(1)}(A_\mu^\alpha,j_\mu)}
\delta\left(\partial_\tau A^\alpha_\tau+
L\partial_iA^\alpha_i
-L^{1/2}\varphi
\right)\nonumber\\
&=\det\left(\partial_\tau^2+L\nabla^2\right)
\int\mathcal D\alpha
\int \mathcal D A^\alpha_\mu\,
e^{-\int_X\mathcal L_{ U(1)}(A^\alpha_\mu,j_\mu)}
e^{-
\int_X\frac{\left(
L^{-1/2}
\partial_\tau A^\alpha_\tau+
L^{1/2}\partial_iA^\alpha_i
\right)^2}{2\xi}}.
\end{align}
It is clear in the second line that $\alpha$ has become a dummy parameter and naturally
the functional integral
over $\alpha$ factors out. The procedure achieves its
goal of summing over unique gauge field configurations, and with the Feynman-'t~Hooft gauge 
choice $\xi=1$, further simplifies the gauge fixed action,
\beq
\mathcal L_A
&=\frac{1}{2}F_{\tau i}F_{\tau i}
+\frac{1}{4}F_{ij}LF_{ij}
+\frac{1}{2}\left(
L^{-1/2}\partial_\tau A_\tau
+L^{1/2}\partial_i A_i
\right)^2,
\eeq
which is no longer singular and yields a simple, diagonal photon propagator~\cite{anselmi10}:
\begin{align}
\langle A_\mu(-K)A_\nu(K)\rangle=
        \delta_{\mu\tau}\delta_{\nu\tau}
        \frac{L(k)}{\omega^2+k^2 L(k)}
        +
        \delta_{\mu i}\delta_{\nu j}
        \frac{\delta_{ij}}{\omega^2+k^2L(k)},
\end{align}
where $L(k)=c^2+v^2k^2$ (see also Fig.~\ref{fig:feynman2}).

\subsection{Zeeman coupling}
\label{app:Zeeman}

In $d=3$, the first-quantized Zeeman Hamiltonian for spin-3/2 fermions can be written as
\begin{align}\label{zeeman}
H_Z=-g\b{B}\cdot\b{J}=-\frac{g}{2}\Sigma^{ij}F_{ij},
\end{align}
where the spin-3/2 matrices $\b{J}=(J_x,J_y,J_z)$ are generators of $SO(3)$ acting on 4-component spinors. Those generators can also be packaged as the antisymmetric rank-2 tensor $\Sigma^{ij}=\epsilon_{ijk}J_k$ which, when contracted with the spatial part $F_{ij}$ of the field strength tensor, forms an $SO(3)$ invariant object. Indeed, the spin-3/2 rotation operator $U(R)$ is written for infinitesimal rotations as:
\begin{align}\label{UR}
U_{\alpha\beta}(R)=\delta_{\alpha\beta}-\frac{i}{2}\theta_{ij}[\Sigma^{ij}]_{\alpha\beta},
\end{align}
where $\theta_{ij}=-\theta_{ji}\equiv\theta\epsilon_{ijk}\hat{n}_k$, for a rotation by infinitesimal angle $\theta$ about axis $\hat{\b{n}}$. In general, under an $SO(3)$ rotation, $\Sigma^{ij}$ transforms as:
\begin{align}
U^\dag(R)\Sigma^{ij}U(R) = R_{ik}\Sigma^{kl}R^T_{lj},
\end{align}
where $R_{ij}$ is an $SO(3)$ rotation matrix and $U_{\alpha\beta}(R)$ is the corresponding unitary transformation in the space of 4-dimensional spinors.

To generalize Eq.~(\ref{zeeman}) to $d=4$ dimensions, as required by our RG analysis, we seek an object $\Sigma^{ij}$ which transforms as an antisymmetric rank-2 tensor under $SO(4)$ rotations in the 16-dimensional spinor space to which $d=4$ Luttinger fermions belong~\cite{janssen15}. This object can then be contracted with the $d=4$ field strength tensor $F_{ij}$ to form an $SO(4)$-invariant term in the Hamiltonian. We simply need to find those matrices $\Sigma^{ij}$ in Eq.~(\ref{UR}) such that the object $G_{ij}$ in the QBC Hamiltonian $H=-G_{ij}p_ip_j$~\cite{janssen15},
\begin{align}
G_{ij}=\sqrt\frac{d}{2(d-1)}\Lambda^a_{ij}\Gamma_a,
\end{align}
transforms as a rank-2 tensor under $SO(4)$ rotations in the 16-dimensional spinor space:
\begin{align}
U^\dag(R)G_{ij}U(R)=R_{ik}G_{kl}R^T_{lj}.
\end{align}
For an infinitesimal rotation, 
$R_{ij}=\delta_{ij}-(i/2)\theta_{kl}[\mathcal J^{kl}]_{ij}$,
the construction yields, to $\c{O}(\theta)$,
\beq
G_{ij}-\frac{i}{2}\theta_{uv}[G_{ij},\Sigma^{uv}]
&=
G_{ij}-\frac{i}{2}\theta_{uv}
\left(
[\mathcal J^{uv}]_{ik}\delta_{jl}
+\delta_{ik}[\mathcal J^{uv}]_{jl}
\right)
G_{kl}.
\eeq
The 16-dimensional spinor space is spanned by the matrices 
$1$, $\Gamma^a$, and $\Gamma^{ab}$, so we can decompose $\Sigma^{uv}=A^{uv}+B^{uv}_a\Gamma^a
+C^{uv}_{ab}\Gamma^{ab}
$. In order to derive these $A$, $B$, and $C$ symbols, we compute the commutator on the left-hand side and find that,
\beq
[G_{ij},\Sigma^{uv}]=-(2\sqrt 2i)\sqrt\frac{d}{d-1}
\left[C^{uv}_{ab}\Lambda^b_{ij}\Gamma^a+B^{uv}_a\Lambda^{ij}_e\Gamma^{ea}\right].
\eeq
It is clear that for $\Sigma^{uv}$ to generate $SO(4)$ rotations, the symbols $B^{uv}_a$ must identically 
vanish. For the right-hand side, we have the expression,
\beq
\left(
[\mathcal J^{uv}]_{ik}\delta_{jl}
+\delta_{ik}[\mathcal J^{uv}]_{jl}
\right)
G_{kl}=-i\sqrt\frac{d}{2(d-1)}\Gamma^e\left(\delta_{iv}\Lambda^e_{ju}
-\delta_{iu}\Lambda^e_{vj}+\Lambda^e_{iu}\delta_{jv}-\Lambda^e_{iv}\delta_{ju}\right).
\eeq
Comparing both sides, we have the relation for the $C$ symbols,
\beq
4C^{uv}_{ab}\Lambda^b_{ij}
=\left(\delta_{iv}\Lambda^a_{ju}
-\delta_{iu}\Lambda^a_{vj}+\Lambda^a_{iu}\delta_{jv}-\Lambda^a_{iv}\delta_{ju}\right),
\eeq
from which the antisymmetry of the symbol under $u\leftrightarrow v$ is evident, as expected. To
enumerate the symbol, we contract both sides with $\Lambda^c_{ji}$, and use the property 
$\tr(\Lambda^a\Lambda^b)=2\delta^{ab}$~\cite{janssen15} to obtain,
\beq
C^{uv}_{ab}=\frac{1}{4}\left(
\Lambda^a_{ui}\Lambda^b_{iv}
-
\Lambda^a_{vi}\Lambda^b_{iu}
\right)
=\frac{1}{4}[\Lambda^a,\Lambda^b]_{uv},
\eeq
which completes our mapping of the $SO(4)$ rotation generator in this 16-dimensional
spinor space:
\beq\label{Sigmaij}
\Sigma^{ij}&=C^{ij}_{ab}\Gamma^{ab}=\frac{1}{4}[\Lambda_a,\Lambda_b]^{ij}\Gamma^{ab}.
\eeq
Using the explicit expressions for the $3\times 3$ Gell-Mann matrices~\cite{janssen15} and the $4\times 4$ gamma matrices~\cite{murakami04}, we verify explicitly that Eq.~(\ref{Sigmaij}) gives $\Sigma^{ij}=\epsilon_{ijk}J_k$ in $d=3$ where $J_k$ are the usual spin-3/2 matrices. The $d=4$ Zeeman Hamiltonian thus again has the form $H_Z=-\frac{g}{2}\Sigma^{ij}F_{ij}$ as in Eq.~(\ref{zeeman}), but with $\Sigma^{ij}$ given in Eq.~(\ref{Sigmaij}).

\subsection{Particle-hole transformation}
\label{app:PHS}

Under the substitution (\ref{PHS}), the temporal term in the fermion Lagrangian (\ref{Lpsi}) transforms as:
\begin{align}
\psi_\alpha^\dag(\partial_\tau+ieA_\tau)\psi_\alpha&\longrightarrow\psi_\alpha^T(\partial_\tau-ieA_\tau)(\psi_\alpha^\dag)^T
=-(\partial_\tau\psi_\alpha^\dag)\psi_\alpha
+\psi_\alpha^\dag ieA_\tau\psi_\alpha=\psi_\alpha^\dag(\partial_\tau+ieA_\tau)\psi_\alpha,
\end{align}
using the anticommutation property of Grassmann variables and integration by parts. Likewise, for the spatial term we have:
\begin{align}\label{spatialPHS}
\psi_\alpha^\dag\Lambda_a^{ij}\Gamma^aD_iD_j\psi_\alpha\longrightarrow\psi_\alpha^T\Lambda_a^{ij}(-\Gamma^{a*})(\partial_i-ieA_i)(\partial_j-ieA_j)(\psi_\alpha^\dag)^T
&=-\psi_\alpha^\dag\Lambda_a^{ij}(-\Gamma^{a*})^T(\partial_j+ieA_j)(\partial_i+ieA_i)\psi_\alpha\nn\\
&=\psi_\alpha^\dag\Lambda_a^{ij}\Gamma^aD_iD_j\psi_\alpha,
\end{align}
using in addition the fact that the $\Gamma^a$ are Hermitian and the $\Lambda_a$ are symmetric. Thus the Lagrangian (\ref{Lpsi}) is even under the particle-hole transformation. However, the Zeeman term (\ref{Lint}) involves the matrices $\Gamma^{ab}$ which, under the substitution $\Gamma^a\rightarrow\Gamma^{a*}$, transform as:
\begin{align}
\Gamma^{ab}=\frac{1}{2i}[\Gamma^a,\Gamma^b]\longrightarrow\frac{1}{2i}[-\Gamma^{a*},-\Gamma^{b*}]=-\left(\frac{1}{2i}[\Gamma^a,\Gamma^b]\right)^*=-\Gamma^{ab*}.
\end{align}
Therefore, under the particle-hole substitution (\ref{PHS}), the Zeeman term (\ref{Lint}) transforms as:
\begin{align}
\psi_\alpha^\dag\Sigma^{ij}\psi_\alpha F_{ij}=\frac{1}{4}\psi_\alpha^\dag[\Lambda_a,\Lambda_b]^{ij}\Gamma^{ab}\psi_\alpha F_{ij}\longrightarrow\frac{1}{4}\psi_\alpha^T[\Lambda_a,\Lambda_b]^{ij}(-\Gamma^{ab*})(\psi_\alpha^\dag)^T(-F_{ij})&=-\frac{1}{4}\psi_\alpha^\dag[\Lambda_a,\Lambda_b]^{ij}(\Gamma^{ab*})^T\psi_\alpha F_{ij}\nn\\
&=-\psi_\alpha^\dag\Sigma^{ij}\psi_\alpha F_{ij},
\end{align}
using the fact that the $\Gamma^{ab}$ are Hermitian, since the $\Gamma^a$ are, and that $F_{ij}\rightarrow-F_{ij}$ since $A_i\rightarrow-A_i$. Finally, following similar steps as in Eq.~(\ref{spatialPHS}), we find that a gauge-invariant quadratic term
\begin{align}
\psi_\alpha^\dag\b{D}^2\psi_\alpha\longrightarrow
\psi_\alpha^T\delta_{ij}(\partial_i-ieA_i)(\partial_j-ieA_j)(\psi_\alpha^\dag)^T=-\psi_\alpha^\dag\delta_{ij}(\partial_j+ieA_j)(\partial_i+ieA_i)\psi_\alpha=-\psi_\alpha^\dag\b{D}^2\psi_\alpha,
\end{align}
is also odd under the particle-hole transformation.

\section{Diagrammatic perturbation theory}
\label{rg_full}

\begin{figure}[t]
  \includegraphics[width=0.85\columnwidth]{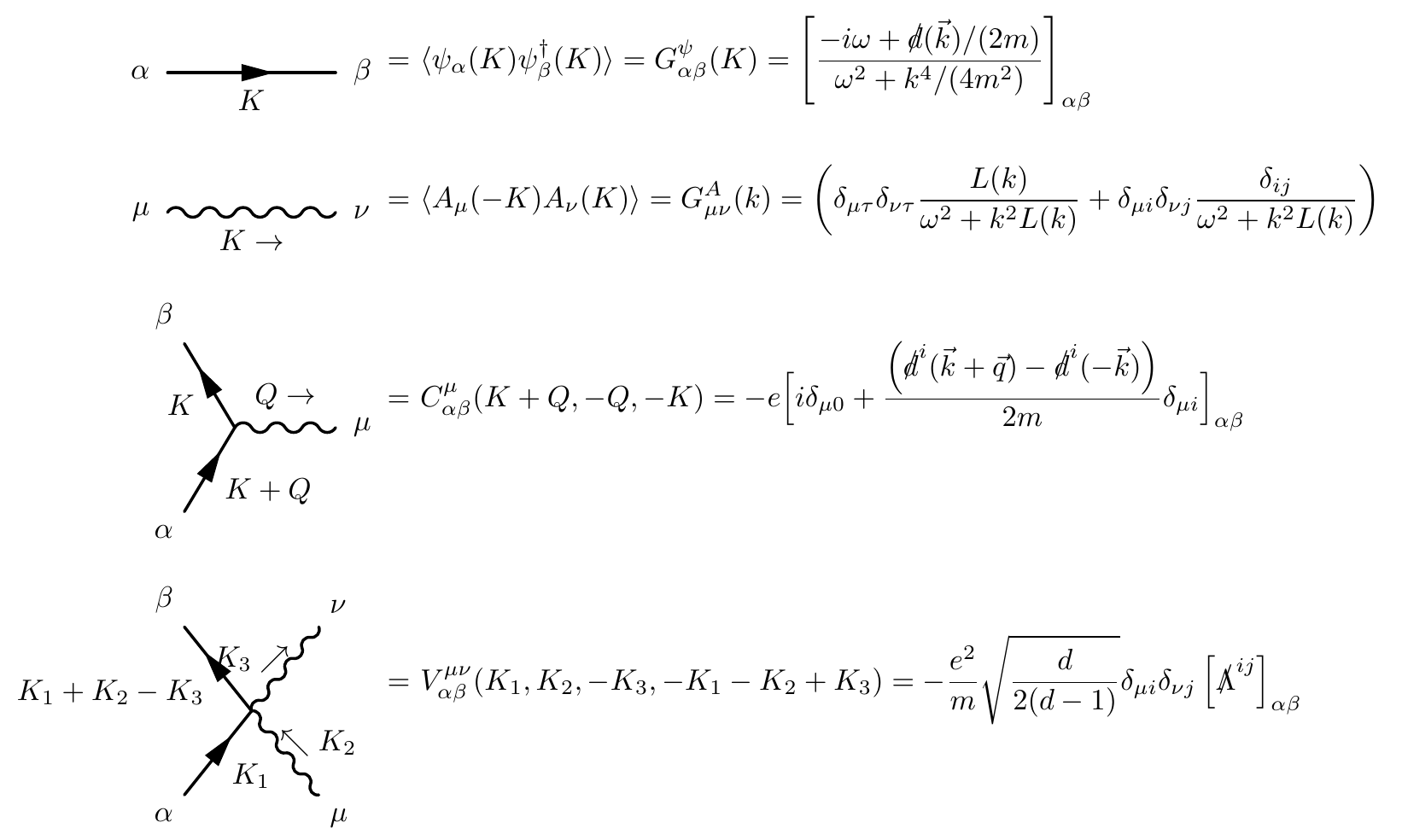}
   \caption{Feynman rules for the $U(1)$ gauge theory of Luttinger QBC fermions. Solid line: fermion propagator $G^\psi$; wavy line: photon propagator $G^A$; three-point vertex: QED vertex $C^\mu\propto e$; four-point vertex: QED seagull vertex $V^{\mu\nu}\propto e^2$.}
   \label{fig:feynman2}
\end{figure}

In this section, we give further details concerning the diagrammatic evaluation of the renormalization constants $\gamma_i=Z_i-1$. Adopting the Fourier transformation convention for the fields,
\begin{align}
A_\mu(X)=\int\frac{d^{d+1}K}{(2\pi)^{d+1}}A_\mu(K)e^{iK\cdot X},\hspace{10mm}
\psi(X)=\int\frac{d^{d+1}K}{(2\pi)^{d+1}}\psi(K)e^{iK\cdot X},
\end{align}
we set up the diagrammatic rules for a loop-perturbation expansion of
the minimal interacting theory of the charged Luttinger fermions, Eq.~\eqref{L}.
As a consistent choice of notation, momentum entering a Feynman diagram
vertex is considered positive, while the momentum leaving a 
vertex is considered negative. The momentum-space
Feynman rules for our model action are given in Fig.~\ref{fig:feynman2}. The last vertex (QED seagull vertex) already contains the symmetry factor of permutation between photonic legs. 

Due to the gauge symmetry, the fermion-photon QED vertex does not 
receive independent renormalization, and the one-loop structure of the renormalized
action emerges from the self-energy corrections alone. Using the one-particle 
irreducible (1PI) diagrams presented in the main text (Fig.~\ref{pert_theory}), the one-loop self-energy corrections
evaluate to:
\beq
\Sigma_{\rm 1L}^\psi(K)
=& 
\int\frac{d^{d+1}Q}{(2\pi)^{d+1}} C^{\mu}(Q,-Q+K,-K)G^\psi(Q)
C^{\nu}(K,Q-K,-Q)G^{A}_{\mu\nu}(Q-K)\nonumber\\
&+
\int\frac{d^{d+1}Q}{(2\pi)^{d+1}} V^{\mu\nu}(K,Q,-Q,-K)G^{A}_{\mu\nu}(Q),\label{SigmaPsi}\\
\Sigma^A_{\mu\nu,\rm 1L}(K) 
=&-
\int\frac{d^{d+1}Q}{(2\pi)^{d+1}} \tr\left[
  C_\mu(Q+K,-K,Q)G^\psi(Q+K)
  C_\nu(Q,K,-Q-K)G^\psi(Q)
  \right]\nonumber\\
&-
\int\frac{d^{d+1}Q}{(2\pi)^{d+1}} \tr\left[V^{\mu\nu}(Q,K,-K,-Q)G(Q)\right],
\label{SigmaA}
\eeq
where $Q=(\omega,\b{q})$ is the internal frequency-momentum. To integrate out high-energy 
fermionic and photonic modes,
 the internal momenta in these loop integrals
are restricted within the momentum shell 
$|\b{q}|\in (\Lambda/b,\Lambda)$. The frequency integrals within the loops are
all convergent and they are computed exactly. The remaining momentum integrals 
for the self-energy corrections lead to complicated expressions. 
To extract the renormalization-scale ($b$) dependent logarithmic factors of the momentum integrals
in $d=4$, we expand the self-energy matrices
in powers of small external frequency and momentum,
\beq
\Sigma_{\rm 1L}(K)=\Sigma_{\rm 1L}(0)
+K^\mu\partial_\mu\Sigma_{\rm 1L}(K)|_{K\rightarrow 0}
+\frac{K^\mu K^\nu}{2}\partial_\mu\partial_\nu\Sigma_{\rm 1L}(K)|_{K\rightarrow 0}
+\dots
\eeq
Using the identities for the Gell-Mann matrices \eqref{lambda_id} after taking spinor traces,
the expansion coefficients only involve internal momenta in the form of generic
integrals
$
\int d^dq/(2\pi)^d f(q^2) q_i q_j q_k\dots
$, that can be reduced further using well-known tensor-integral identities, such as:
\beq
\int\frac{d^dq}{(2\pi)^d}f(q^2)q_iq_j
&=\frac{\delta_{ij}}{d}
\int\frac{d^dq}{(2\pi)^d}f(q^2)q^2,\lb
\int\frac{d^dq}{(2\pi)^d}f(q^2)q_iq_j q_k q_l
&=\frac{
  \delta_{ij}\delta_{kl}
  +\delta_{ik}\delta_{jl}
  +\delta_{il}\delta_{jk}
}{d(d+2)}
\int\frac{d^dq}{(2\pi)^d}f(q^2)q^4,
\eeq
and similarly for higher-rank tensors. In such types of integral 
the only nonzero contribution comes from cases where all the internal momentum
indices can be pairwise contracted. The contraction identities can be proved with 
the help of the Laplace transform of the (well-behaved) 
function $f(q^2)$. The Laplace transform of integrals of the type,
\beq
K_{ijkl}&=\int\frac{d^dq}{(2\pi)^d}f(q^2)q_i q_j q_k q_l=\int^\infty_0dt\int\frac{d^dq}{(2\pi)^d}
\tilde f(t)e^{-q^2t}q_iq_jq_kq_l,
\eeq
can be recast in terms of a generating variable $Y_i$ such that,
\beq
K_{ijkl}
&=
\left.
\frac{\partial^4}{\partial Y_i
\partial Y_j
\partial Y_k
\partial Y_l
}
\left[
\int^\infty_0dt\int\frac{d^dq}{(2\pi)^d}
\tilde f(t)e^{-q^2t+Y_iq_i}
\right]
\right|_{Y_i\rightarrow 0}
,\nonumber\\
&=
\left.
\frac{\partial^4}{
  \partial Y_i
  \partial Y_j
  \partial Y_k
  \partial Y_l
}
\left[
\int^\infty_0dt\tilde f(t)
\frac{
  e^{Y_i^2/(4t)}
}{(2\pi)^{d/2}(2t)^{d/2}}
\right]
\right|_{Y_i\rightarrow 0}
,\nonumber\\
&=
\int^\infty_0dt
\frac{
\tilde f(t)
}{(2\pi)^{d/2}(2t)^{d/2}}
\left(
\frac{\delta_{ij}}{2t}
\frac{\delta_{kl}}{2t}
+
\frac{\delta_{ik}}{2t}
\frac{\delta_{jl}}{2t}
+
\frac{\delta_{il}}{2t}
\frac{\delta_{jk}}{2t}
\right).
\eeq
On the other hand it follows that,
\beq
\int\frac{d^dq}{(2\pi)^d}f(q^2)q^4
&=\int^\infty_0 dt\int\frac{d^dq}{(2\pi)^d}
\tilde f(t)e^{-q^2t}q^4,\nonumber\\
&=\int^\infty_0 dt
\tilde f(t)
\frac{\partial^2}{\partial^2 t}
\int\frac{d^dq}{(2\pi)^d}e^{-q^2t},\nonumber\\
&=\int^\infty_0 dt
\tilde f(t)
\frac{\partial^2}{\partial^2 t}
\left(
\frac{1}{(2\pi)^{d/2}(2t)^{d/2}}\right),\nonumber\\
&=\int^\infty_0 dt
\frac{\tilde f(t)}{(2\pi)^{d/2}(2t)^{d/2}}
\left(\frac{d(d+2)}{4t^2}\right).
\eeq
Combining these two results we find that,
\beq
K_{ijkl}&=\int\frac{d^dq}{(2\pi)^d}f(q^2)q_i q_j q_k q_l
=\frac{1}{d(d+2)}
\int\frac{d^dq}{(2\pi)^d}f(q^2)q^4
\left(
\delta_{ij}
\delta_{kl}
+
\delta_{ik}
\delta_{jl}
+
\delta_{il}
\delta_{jk}
\right),
\eeq
and similarly,
\beq
K_{ijklmn}&=\int\frac{d^dq}{(2\pi)^d}f(q^2)q_i q_j q_k q_l q_m q_n
=\frac{1}{d(d+2)(d+4)}
\int\frac{d^dq}{(2\pi)^d}f(q^2)q^6
\left(
\delta_{ij}
\delta_{kl}\delta_{mn}
+
\dots
\right),
\eeq
and so on.

Applying these algebraic manipulations to Eq.~(\ref{SigmaPsi}),
we obtain the following 
one-loop self-energy correction to the fermion propagator [Fig.~\ref{pert_theory}(a)]:
\beq 
\Sigma^\psi_{1L}(K)&= -\gamma_1\left(i\omega\right)
-\gamma_2 \frac{\slashed d(\vec k)}{2m},
\eeq 
with the renormalized inverse propagator now given by
$(\tilde G^{\psi})^{-1}=(G^{\psi})^{-1}-\Sigma^\psi_{1L}$. This leads to
the effective one-loop renormalized fermionic action in Eq.~(\ref{Leff}). The
term involving the quartic vertex in Eq.~\eqref{SigmaPsi}
vanishes and does not contribute to the fermionic self-energy. The
two renormalization constants are given by:
\begin{align}
\gamma_1&=
   \frac{\alpha  (4 K+4 \rho -3) \left(4 \sqrt{K+\rho }-4 K-4 \rho -1\right)}{2 \sqrt{K+\rho } (4 K+4 \rho -1)^2},\nn\\
   \gamma_2
   &=
   \frac{\alpha}{36 (K+\rho )^{3/2} (4 K+4 \rho -1)^3}
\Bigg[  \Big(48 K^4 \Big(16 \sqrt{K+\rho }-15\Big)+8 K^3 \Big(12 \rho  \Big(32 \sqrt{K+\rho }-33\Big)+16 \sqrt{K+\rho }+3\Big)\lb
&\phantom{=}+K^2 \Big(8 \rho
    \Big(576 \rho  \sqrt{K+\rho }+112 \sqrt{K+\rho }-666 \rho -21\Big)+16 \sqrt{K+\rho }-13\Big)\lb
    &\phantom{=}+4 K \rho  \Big(3 \Big(8 \sqrt{K+\rho }-5\Big)+4 \rho 
   \Big(12 \rho  \Big(16 \sqrt{K+\rho }-21\Big)+112 \sqrt{K+\rho }-39\Big)\Big)\lb
   &\phantom{=}+\rho  \Big(16 \rho  \Big(\rho  \Big(24 \rho  \Big(2 \sqrt{K+\rho
   }-3\Big)+64 \sqrt{K+\rho }-27\Big)+5 \sqrt{K+\rho }-2\Big)+5\Big)+K\Big)\Bigg].
\end{align}

The photon self-energy (\ref{SigmaA}) is computed in a similar manner and the
four-point vertex once again offers no contribution. The three renormalization
constants from the photon self-energy correction [Fig.~\ref{pert_theory}(b)] entering the renormalized
effective action in Eq.~\eqref{Leff} are as follows:
\beq
\gamma_3=2 \alpha  N_f,\hspace{10mm}
\gamma_4=\frac{13 \alpha  N_f}{36 K},\hspace{10mm}
\gamma_5=-\frac{5 \alpha  N_f}{27 \rho }.
\eeq
Note that the dependence $\propto 1/K$ in $\gamma_4$ is what enables the RG equation (\ref{dKdl}) to have a structure similar to the Wilson-Fisher equation (\ref{WF}).

Finally, the dynamic critical exponent (\ref{Modifiedz}) of the theory has the following expression in terms of the renormalized couplings:
\beq
z-2&=\frac{\alpha}{36
   (K+\rho )^{3/2} (4 K+4 \rho -1)^3}\bigg[\big(-48 K^4 \big(16 \sqrt{K+\rho }+9\big)+8 K^3
   \big(-12 \rho  \big(32 \sqrt{K+\rho }+15\big)+128
   \sqrt{K+\rho }+105\big)\lb
   &\phantom{=}+K^2 \big(-144 \rho ^2 \big(32
   \sqrt{K+\rho }+11\big)+40 \rho  \big(64 \sqrt{K+\rho
   }+69\big)-1168 \sqrt{K+\rho }+85\big)\lb
   &\phantom{=}+K \big(-4 \rho 
   \big(4 \rho  \big(12 \rho  \big(16 \sqrt{K+\rho }+3\big)-104
   \sqrt{K+\rho }-201\big)+600 \sqrt{K+\rho }-51\big)+216
   \sqrt{K+\rho }-55\big)\lb
   &\phantom{=}+\rho  \big(-8 \rho  \big(2 \rho 
   \big(48 \rho  \sqrt{K+\rho }-8 \sqrt{K+\rho }-81\big)+154
   \sqrt{K+\rho }-13\big)+216 \sqrt{K+\rho }-59\big)\big)\bigg].
   \label{dynx_formula}
\eeq

\section{Stability of Lifshitz-QED fixed point}
\label{fixeds}

The two functions appearing in the RG equations (\ref{dadl}-\ref{drhodl}) are defined as:
\beq
f_1(\rho)&=
\frac{4 \rho  \left(4 \sqrt{\rho } \left(12 \left(\rho +\sqrt{\rho }\right)+7\right)-65\right)+59+20\sqrt{\rho }}{36\sqrt{\rho } \left(2 \sqrt{\rho }-1\right)
   \left(2 \sqrt{\rho }+1\right)^3}
,\lb
f_2(\rho,N_f)&=
\frac{1}{108} \left(-4 (54 \rho +5) N_f-\frac{6 \sqrt{\rho } \left(4 \sqrt{\rho } \left(48 \rho ^{3/2}+48 \rho ^2+28 \rho -65 \sqrt{\rho
   }+5\right)+59\right)}{\left(2 \sqrt{\rho }-1\right) \left(2 \sqrt{\rho }+1\right)^3}\right)
.
\eeq
With the one-loop RG equations (\ref{dadl}-\ref{drhodl}), for $N_f\geq 2$ we find a Lifshitz-QED (LQED)
multicritical point with finite charge coupling and finite photon stiffness. For $N_f=2$, the magnitudes of the critical couplings are numerically found to be:
\beq
&\alpha^*=0.958584\epsilon, \ \ 
K^*=-0.346155\epsilon, \ \ 
\rho^*=0.193547.
\eeq
The stability matrix $M_{ij}
=\partial\beta_i/\partial\lambda_j$, with couplings $\lambda_j\in\{\alpha,K,\rho\}$ 
and beta functions $\beta_i=d\lambda_i/d\ell$,
can be enumerated at the fixed point to determine its stability. The 
stability matrix up to linear order in $\epsilon$ is found to be:
\beq
M_{\rm LQED}=\begin{pmatrix}
  -\epsilon&&0&&0\\
  0.722222-0.662399\epsilon&&2+1.83434\epsilon&&0\\
  -2.92933\epsilon&&8.11199\epsilon&&14.3677\epsilon
\end{pmatrix}.
\label{lqed_stab}
\eeq
The eigenvalues, found on the diagonal, are:
\begin{align}
-\epsilon,\hspace{10mm}
2+1.83434\epsilon,\hspace{10mm}
14.3677\epsilon,
\end{align}
to $\c{O}(\epsilon)$. Since there exists more than one unstable direction, the fixed point is multicritical. The same number of stable/unstable directions is found for other values of $N_f\geq 2$.

\twocolumngrid

\bibliography{lqed}

\begin{thebibliography}{72}
\bibitem [{LAB()}]{LABnote}%
  \BibitemOpen
  \href@noop {} {}\bibinfo {note} {The precise values of $\alpha_*$ and $z$ at
  the fixed point differ from those given in Ref.~\cite{moon13}, because unlike
  there, we compute the renormalization constants at the upper-critical
  dimension $d=4$, whereas they compute everything in the spatial dimension
  $d=3$ except radial momentum integrals which are computed in $d=4$ to capture
  the logarithmic divergence. Our computation reproduces their result once
  their renormalization prescription is adopted. The $d=4$ prescription has the
  advantage over the $d=3$ prescription that it maintains invariance under
  shifts of integration variable in loop-momentum integrals~\cite{boettcher17},
  but the qualitative physics is independent of the choice of
  prescription.}\BibitemShut {Stop}%
\end{thebibliography}

\end{document}